\documentclass[conference]{IEEEtran}

\def\BibTeX{{\rm B\kern-.05em{\sc i\kern-.025em b}\kern-.08emT\kern-.1667em\lower.7ex\hbox{E}\kern-.125emX}}
    
\usepackage{nicefrac}
\usepackage{siunitx}
\usepackage{array,framed}
\usepackage{booktabs}
\usepackage{
  color,
  float,
  epsfig,
  wrapfig,
  graphics,
  graphicx,
  subcaption
}
\usepackage{textcomp,amssymb}
\usepackage{setspace}
\usepackage{latexsym,fancyhdr,url}
\usepackage{enumerate}
\usepackage{algorithm2e}
\usepackage{algpseudocode}
\usepackage{graphics}
\usepackage{xparse} %
\usepackage{xspace}
\usepackage{multirow}
\usepackage{csvsimple}
\usepackage{balance}
\usepackage{diagbox}
\usepackage{bbding}
\usepackage{colortbl}
\usepackage{enumitem}
\usepackage{xurl}

\definecolor{lightlightgray}{RGB}{220,220,220}

\usepackage{
  tikz,
  pgfplots,
  pgfplotstable
}
\usepackage{hyperref}

\usetikzlibrary{
  shapes.geometric,
  arrows,
  external,
  pgfplots.groupplots,
  matrix
}

\pgfplotsset{compat=1.9}

\usepackage{mathtools,}

\DeclareMathAlphabet{\mathcal}{OMS}{cmsy}{m}{n}

\DeclareGraphicsExtensions{%
    .png,.PNG,%
    .pdf,.PDF,%
    .jpg,.mps,.jpeg,.jbig2,.jb2,.JPG,.JPEG,.JBIG2,.JB2}

\DeclareUnicodeCharacter{00A0}{ }

\usepackage{xparse}
\newcommand{\bnm}{\begin{newmath}}
\newcommand{\enm}{\end{newmath}}

\newcommand{\bea}{\begin{eqnarray*}}%
\newcommand{\eea}{\end{eqnarray*}}%

\newcommand{\bne}{\begin{newequation}}
\newcommand{\ene}{\end{newequation}}

\newcommand{\bal}{\begin{newalign}}
\newcommand{\eal}{\end{newalign}}

\newenvironment{newalign}{\begin{align}%
\setlength{\abovedisplayskip}{4pt}%
\setlength{\belowdisplayskip}{4pt}%
\setlength{\abovedisplayshortskip}{6pt}%
\setlength{\belowdisplayshortskip}{6pt} }{\end{align}}

\newenvironment{newmath}{\begin{displaymath}%
\setlength{\abovedisplayskip}{4pt}%
\setlength{\belowdisplayskip}{4pt}%
\setlength{\abovedisplayshortskip}{6pt}%
\setlength{\belowdisplayshortskip}{6pt} }{\end{displaymath}}

\newenvironment{newequation}{\begin{equation}%
\setlength{\abovedisplayskip}{4pt}%
\setlength{\belowdisplayskip}{4pt}%
\setlength{\abovedisplayshortskip}{6pt}%
\setlength{\belowdisplayshortskip}{6pt} }{\end{equation}}

\newcounter{ctr}

\newcounter{mytable}
\def\mytable{\begin{centering}\refstepcounter{mytable}}
\def\endmytable{\end{centering}}

\newcounter{myfig}
\def\myfig{\begin{centering}\refstepcounter{myfig}}
\def\endmyfig{\end{centering}}

\newlength{\saveparindent}
\newlength{\saveparskip}
\newcommand{\E}{{\rm I\kern-.3em E}}

\renewcommand{\eqref}[1]{\mbox{Equation~(\ref{#1})}}

\def \part {part}

\renewcommand{\paragraph}[1]{\vspace*{6pt}\noindent\textbf{#1}\;}

\def \blackslug{\hbox{\hskip 1pt \vrule width 4pt height 8pt
    depth 1.5pt \hskip 1pt}}
\def \qed{\quad\blackslug\lower 8.5pt\null\par}

\newcounter{mynote}[section]

\newcommand\ignore[1]{}

\newcounter{rcnote}[section]

\newcounter{mrnote}[section]

\newcounter{fknote}[section]

\newcounter{anote}[section]

\DeclareMathSymbol{\mlq}{\mathord}{operators}{``}
\DeclareMathSymbol{\mrq}{\mathord}{operators}{`'}

\newcommand{\rhf}[2]{R_{f, \gamma}}

\DeclareDocumentCommand{\edist}{o o}{
  \ensuremath{
    \IfNoValueTF{#1}{{d}}{{\sf d}(#1,#2)}
  }
}

\newcommand{\olrk}[1]{\ifx\nursymbol#1\else\!\!\mskip4.5mu plus 0.5mu\left(\mskip0.5mu plus0.5mu #1\mskip1.5mu plus0.5mu \right)\fi}

\NewDocumentCommand{\indseq}{ O{1} O{r} }{{#1}\ldots {#2}}

\iftrue

\newcommand{\nsection}[1]{\section{#1}}
\newcommand{\nsubsection}[1]{\subsection{#1}}
\newcommand{\nsubsubsection}[1]{\subsubsection{#1}}

\else

\newcommand{\nsection}[1]{\vspace{-0.05in}\section{#1}\vspace{-0.0in}}
\newcommand{\nsubsection}[1]{\vspace{-0.05in}\subsection{#1}\vspace{-0.05in}}
\newcommand{\nsubsubsection}[1]{\vspace{-0.0in}\subsubsection{#1}\vspace{-0.0in}}

\fi

\newcommand{\cut}[1]{}

\makeatletter
\newcommand\notsotiny{\@setfontsize\notsotiny\@vipt\@viipt}
\makeatother

\usepackage{cite}

\ifCLASSINFOpdf
\else
\fi

\begin{document}
\fancyhf{} %

\title{Invisible Reflections: Leveraging Infrared Laser Reflections to Target Traffic Sign Perception}

\author{\IEEEauthorblockN{
Takami Sato\IEEEauthorrefmark{1}\IEEEauthorrefmark{2}\thanks{\IEEEauthorrefmark{1} denotes co-first authors.},
Sri Hrushikesh Varma Bhupathiraju\IEEEauthorrefmark{1}\IEEEauthorrefmark{3},
Michael Clifford\IEEEauthorrefmark{4},\\ 
Takeshi Sugawara\IEEEauthorrefmark{5},
Qi Alfred Chen\IEEEauthorrefmark{2}, 
Sara Rampazzi\IEEEauthorrefmark{3} 
}
\IEEEauthorblockA{\IEEEauthorrefmark{2}University of California, Irvine; \IEEEauthorrefmark{3}University of Florida;
\IEEEauthorrefmark{4}Toyota InfoTech Labs;
\IEEEauthorrefmark{5}The University of Electro-Communications
}
}

\IEEEoverridecommandlockouts
\makeatletter\def\@IEEEpubidpullup{3\baselineskip}\makeatother
\IEEEpubid{\parbox{\columnwidth}{
    Network and Distributed System Security (NDSS) Symposium 2024\\
    26 February - 1 March 2024, San Diego, CA, USA\\
    ISBN 1-891562-93-2\\
    https://dx.doi.org/10.14722/ndss.2024.231053\\
    www.ndss-symposium.org
}
\hspace{\columnsep}\makebox[\columnwidth]{}}

\maketitle

\begin{abstract}\label{sec:abstract}
All vehicles must follow the rules that govern traffic behavior, regardless of whether the vehicles are human-driven or Connected Autonomous Vehicles (CAVs). Road signs indicate locally active rules, such as speed limits and requirements to yield or stop. Recent research has demonstrated attacks, such as adding stickers or projected colored patches to signs, that cause CAV misinterpretation, resulting in potential safety issues. Humans can see and potentially defend against these attacks. But humans can not detect what they can not observe. We have developed an effective physical-world attack that leverages the sensitivity of filterless image sensors and the properties of Infrared Laser Reflections (ILRs), which are invisible to humans. The attack is designed to affect CAV cameras and perception, undermining traffic sign recognition by inducing misclassification.
In this work, we formulate the threat model and requirements for an ILR-based traffic sign perception attack to succeed. We evaluate the effectiveness of the ILR attack with real-world experiments against two major traffic sign recognition architectures on four IR-sensitive cameras. Our black-box optimization methodology allows the attack to achieve up to a 100\% attack success rate in indoor, static scenarios and a $\geq$80.5\% attack success rate in our outdoor, moving vehicle scenarios. 
We find the latest state-of-the-art certifiable defense is ineffective against ILR attacks as it mis-certifies $\geq$33.5\% of cases. To address this, we propose a detection strategy based on the physical properties of IR laser reflections which can detect 96\% of ILR attacks.

\end{abstract}

\nsection{Introduction} \label{sec:intro}

Every vehicle, whether a connected, autonomous vehicle (CAV), semi-autonomous, or human-driven vehicle, must obey traffic signs.  Disobeying signs can cause potential accidents and threaten human life.
Recent studies~\cite{eykholt2018physical, zhao2018seeing, lovisotto2021slap, jia2022fooling, zhong2022shadows, duan2021adversarial, nassi2020phantom} on traffic sign recognition systems show how physical, adversarial attacks can degrade recognition accuracy.  Such attacks include projecting shadows~\cite{zhong2022shadows},  projecting visible colored patterns~\cite{lovisotto2021slap, duan2021adversarial, nassi2020phantom, yufeng2023light}, and adding stickers to traffic signs~\cite{eykholt2018physical, zhao2018seeing, jia2022fooling}, to induce misclassification.
However, these prior attacks have a clear limitation in stealthiness. Stickers, strong light projections, or shadows inconsistent with the environment are visible to humans, who can detect and mitigate them. For example, in semi-autonomous vehicles, such as Tesla's~\cite{teslamanual}, drivers are required to stay alert and ready to take manual control of the vehicle if needed. As long as visual changes are inevitable, humans may notice them, even if they are subtle.

\begin{figure}[t!]
\begin{center}
    \includegraphics[width=\linewidth]{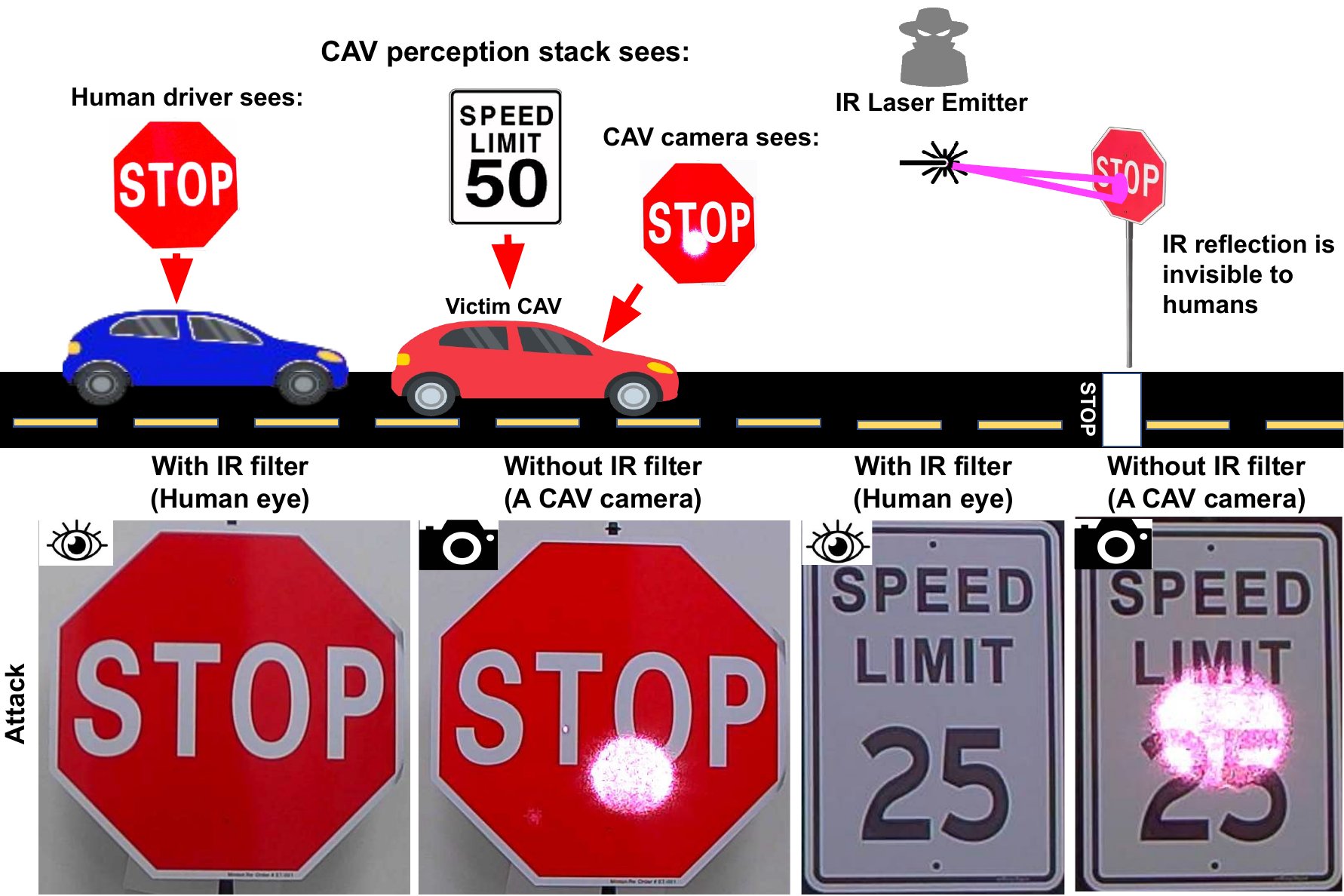}
\end{center}
\caption{Overview of our ILR (Infrared Laser Reflection) attack.  Unlike cameras without IR filters, humans can not see IR light.  When an IR-sensitive camera images an object illuminated by an IR laser, the camera's output is altered at the pixel level.  Our attack causes CAV perception stacks to misclassify traffic signs, causing dangerous misinterpretations.}

\label{fig:demo}
\end{figure}

In this work, we design and demonstrate a novel, invisible attack that is able to mislead traffic sign recognition systems by leveraging patterns of infrared (IR) light reflections that are invisible to humans. As shown in Fig.~\ref{fig:demo}, the reflections are visible to CAV cameras without an IR filter. These cameras capture the attacker's IR light reflections on the target traffic sign and output an altered image that is misinterpreted by the vehicle's perception module in its autonomy stack (e.g., detecting a speed limit sign instead of a stop sign).  

Camera sensors are normally sensitive to photons in both visible and infrared wavelengths. Typically, commercial cameras use IR filters to ensure accurate color reproduction and prevent unwanted contamination by infrared photons. 
However, some autonomous vehicles employ cameras without these filters to improve detection in dark environments~\cite{tesla2020autopilot, mobileye}. Our attack targets sensors lacking these filters.  Moreover, although humans might perceive infrared light reflections through such cameras, CAVs generally do not display the captured images to the drivers, making this attack challenging to detect, or to distinguish from ordinary CAV malfunctions.

For example, the recent I-Can-See-the-Light (ICSL) attack~\cite{wang2021can} takes advantage of filterless sensors by projecting an IR pattern directly onto the camera image sensor in order to create fake objects and induce SLAM errors. Another work describes an invisible mask attack~\cite{zhou2018invisible}, which uses multiple IR emitters inside a hat to evade face recognition surveillance.
However, these attacks only focus on changing traffic light colors and inducing detection errors, leaving unclear the impact of IR-based attacks on CAV traffic sign recognition systems. Furthermore, those attacks either require that the IR light source be aimed continuously and precisely at the target camera on a moving CAV, or only function at short distances.  This limits the attacks' practicality in real-world scenarios. 
These limitations motivate our work.

Our \emph {Infrared Laser Reflection} (ILR) attack causes CAV perception modules to misclassify, or in the worst case misdetect, traffic signs.  We use an IR laser source to reflect IR projections off of a portion of a target sign surface.  Leveraging the unique properties of laser light reflections, an IR-sensitive CAV camera will see the reflected light and incorporate it into the camera's output images.  We call these images \emph{traces}. The vehicle's perception module will then attempt to classify the trace-tainted images from the camera, resulting in misclassification.  Unlike ICSL and similar attacks~\cite{yan2022rolling, jin2023pla}, which require precisely aiming at the target sensor and accurately synchronizing timing, we only need to aim a single IR laser at a target traffic sign. 
The IR laser emitter is static and needs no sophisticated setup to track a moving CAV.
We also find that the laser reflection properties can achieve stable misclassification at long distances with minimum power (up to 25 meters away from the target sign, with a laser power of 26 mW). To maximize the attack effect, we developed a technique for generating optimized traces using the IR laser reflections, which allow the attacker to automatically find the optimal misclassification, minimizing the required laser power, and covering a minimal portion of the target sign with the reflection (7--17\% of real-world stop and speed limit sign size in our outdoor evaluation).

We evaluate our ILR attack's effectiveness against two major traffic sign recognition architectures, using images captured with four different IR-sensitive cameras. Our paper is structured as follows: In~\S\ref{sec:threat_model}, we formulate our threat model. We determine what parameters the attacker can control and what parameters are required to make the attack robust to different conditions. In~\S\ref{sec:methodology}, we describe our attack optimization methodology, which incorporates modeling the attack reflection characteristics to automatically find the optimal location of the projection on the target sign while minimizing the required power and reflection size. In~\S\ref{sec:evaluation}, we demonstrate that our attack achieves up to a 100\% attack success rate in the real world against both stop sign and speed limit sign targets under static, indoor laboratory conditions, and a $\geq$80.5\% attack success rate in outdoor conditions, with a vehicle moving at increasing speeds. In~\S\ref{sec:defense}, we evaluate ILR against the current state-of-the-art certifiable defense PatchCleanser~\cite{xiang2022patchcleanser}, which has been evaluated against patch attacks that target generic image classification tasks for the ImageNet~\cite{deng2009imagenet} and CIFAR-10~\cite{krizhevsky2009learning} datasets.  
We found that PatchCleanser's major intuition does not hold for traffic sign recognition systems, making the defense ineffective against ILR, as PatchCleanser mis-certifies $\geq$33.5\% of attack and benign cases.
To address this limitation, we propose a potential defense strategy based on the unique features of IR laser reflections. Finally, we discuss the limitations of this study in~\S\ref{sec:discussion}.

In summary, our study makes the following contributions:
\begin{itemize}[leftmargin=0.2in]
    \item We identify ILR, a long-distance and human-invisible attack vector that can cause misclassification by traffic sign recognition systems. The ILR attack can not be seen by humans, but can be seen by cameras lacking IR filters. Our attack addresses the aiming and power limitations of previous works by combining invisible laser reflection properties and adversarial optimization.
    
    \item We design a novel methodology to optimize attack effectiveness by simulating IR laser projections, and their  traces on signs, by modeling reflection size, intensity, and position using a black-box optimization.

    \item We evaluate the ILR attack against two different sign recognition architectures using four IR-sensitive cameras.  We confirm that the ILR attack reaches a 100\% attack success rate under indoor laboratory conditions and a  $\geq$80.5\% attack success rate in outdoor, real-world environments with different light conditions and victim vehicle speeds (up to 13 Km/h).
    
    \item We show that the major assumption of the state-of-the-art, certifiable defense, PatchCleanser~\cite{xiang2022patchcleanser}, does not hold in the traffic sign recognition domain. PatchCleanser mis-certifies $\geq$33.5\% of cases. We then propose a potential detection technique based on the unique physical characteristics of the IR laser reflections, which achieves a 96\% True Positive Rate (TPR) and 6.7\% False Positive Rate (FPR) in our proof-of-concept tests.

\end{itemize}

Details and demo videos are available on our website: \textcolor{blue}{\url{https://sites.google.com/view/cav-sec/ilr-attack}}.
\nsection{Background and Related Work} \label{sec:background}

\nsubsection{Vision-Based Traffic Sign Recognition} \label{sec:vision-based} \label{sec:overview_tsd}
Vision-based traffic sign recognition systems use camera sensor outputs as inputs to fast neural networks, which perform real-time object recognition and classification~\cite{Houben-IJCNN-2013,ertler2020mapillary}. These recognition systems have benefits in terms of both capability and cost.  They are also essential, as autonomous vehicles \emph{must} recognize road signs in order to operate safely on public roads. This has driven their wide adoption in autonomous driving systems, such as those offered by OpenPilot~\cite{openpilot} and Tesla~\cite{tesla2020autopilot}.
Ertler~et~al.~\cite{ertler2020mapillary} identify two major vision-based traffic sign recognition architectures: \textit{single-stage} and \textit{two-stage}, as illustrated in Fig.~\ref{fig:single_two_stage_arch}:

\begin{figure}[t!]
\begin{center}
    \includegraphics[width=\linewidth]{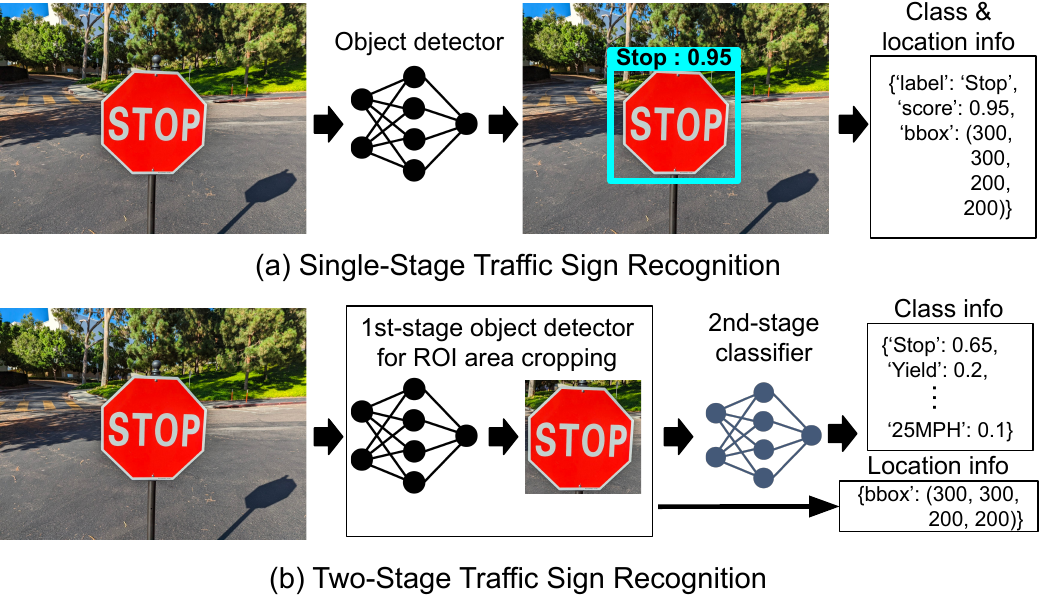}
\end{center}
\vspace{-0.1in}
\caption{The two major  traffic sign recognition system architectures used in our work. (a) A single-stage architecture detects and classifies traffic signs only with a single object detector; (b) A two-stage architecture's first-stage object detector finds and isolates (crops) the traffic sign in the image.  The second-stage classifier provides the sign's class label.
}
\label{fig:single_two_stage_arch}
\end{figure}

\noindent\textbf{Single-Stage Architectures.} Single-stage architectures implement an object detector, such as  YOLO~\cite{yolo9000}, using a multi-class classification head to interpret traffic sign types. While the single-stage architecture is advantageous in terms of computational cost,  Ertler~et~al.~\cite{ertler2020mapillary} report that the single-stage architecture does not yield acceptable performance when the number of classes is large, as with the 314 classes in~\cite{ertler2020mapillary}. Hence, typically, a single-stage architecture is suitable for Level-2 autonomy systems that only need to recognize a limited number of signs.

\noindent\textbf{Two-Stage Architectures.}  A two-stage architecture, which is capable of handling a large number of different signs, uses a first-stage object detector to crop the image to a "Region of Interest" (ROI) that contains the traffic sign.  It then classifies the cropped image in its second stage~\cite{ertler2020mapillary}.  Specifically, the first-stage object detector detects the sign's position, with croping performed regardless of  sign type.  The second stage then classifies the cropped region with a sign type~\cite{chiu2021two}. 

Unfortunately, previous work has shown how real-time sign perception can be vulnerable to attacks that affect what the camera sees~\cite{eykholt2018physical, chen2018shapeshifter, zhao2018seeing}. Our work focuses on a novel perception attack that can affect both one and two-stage architectures.  These are available in production vehicles.

\nsubsection{Human Perception of Visible and IR Light}
\label{ir_color}
Invisible to humans, infrared light is electromagnetic radiation with wavelengths between 780 nm and 1 mm~\cite{someda2017electromagnetic}. The CMOS image sensors used in today's cameras have sensitivity to some IR light.  To match the perception characteristics of human eyes, they usually incorporate a filter that cuts out this IR light~\cite{teledyneImg, IRandvis}. However, to improve camera performance for nighttime driving,  some production CAVs use cameras without IR filters~\cite{Thakur17, tesla2020autopilot, mobileye}. 
Discussion of the prevalence of IR-sensitive cameras is challenging, as manufacturers seldom disclose specific information.  To the best of our knowledge though, Tesla Model 3 cameras lack IR filters~\cite{wang2021can}.

Our attack manipulates what the IR-sensitive camera sensor sees by projecting IR light patterns onto an object in the field of view of a CAV's camera.  Since the unfiltered image sensor in the camera can see IR light, the reflection of the projection becomes part of the sensor's output image. Humans cannot react to the attack since they can not perceive it.

As shown in Fig.~\ref{fig:demo}, the reflection of the IR projection on the sign \emph{is} visible in the output image. This occurs because the sensor can only measure the \emph{intensity} of incident photons at each sensor photosite, not the \emph{wavelength} of each photon. Typically, image sensors have color filters to only allow red, green, or blue photons to hit each photosite~\cite{bayer1976color}. Because this filtering is imperfect, a portion of incident IR light passes through the color filters, reaches the photosites, and is integrated into the output image. Depending on the IR transmittance of these color filters, IR light appears in the output image with a false color, which is usually purple, magenta, or even orange.

\nsubsection{Previous Work and Comparisons} \label{sec:attack_tsd}

Deep Neural Network (DNN) models today are shown to be generally vulnerable to adversarial examples (or adversarial attacks)~\cite{Szegedy2014, goodfellow2014explaining}. These attacks have previously been explored in the physical world~\cite{kurakin2016adversarial_b, sharif2016accessorize, athalye2018synthesizing, brown2017advpatch, chen2018shapeshifter, eykholt2018physical, jack2018caraml, zhao2018seeing, pei2017deepxplore, tian2018deeptest, chernikova2019self, zhou2018deepbillboard, jin2023pla}.
In particular, traffic sign recognition has been shown to be vulnerable to adding small stickers to signs~\cite{eykholt2018physical, chen2018shapeshifter, zhao2018seeing}, visible pattern projection~\cite{lovisotto2021slap, duan2021adversarial, LI2023103345}, and shadow shading~\cite{zhong2022shadows}. 
Unlike physical patch attacks that leave permanent, detectable artifacts (such as small stickers) on the target sign \cite{eykholt2018physical,zhao2018seeing}, our attacks avoid destructive changes or physical alterations to the target through the use of light projection and reflection. More specifically, our attack has the following three major differences or advantages over prior, related work:

\textit{(1) Invisibility and Attacker Capabilities.} As discussed in~\S\ref{ir_color}, our attack is invisible to humans, rendering it more difficult to detect than physical patches~\cite{eykholt2018physical,zhao2018seeing} or visible colored pattern projections~\cite{259721, nassi2020phantom_all, lovisotto2021slap, LI2023103345}. Previous work, such as ICSL~\cite{wang2021can}, remote attacks~\cite{petit2015remote}, and invisible masks~\cite{zhou2018invisible} use IR light to enhance stealthiness.  In contrast to this work, which uses non-coherent IR LED light, the ILR attack uses coherent laser light.  Because coherent light waves are in phase with each other~\cite{svelto2010principles}, laser light remains in a confined, tightly focused, and persistent beam over long distances, with little attenuation. This property can persist even when reflected from a surface, such as that of a sign, allowing us to optimize our projected patterns.  

For example, we demonstrate how our ILR attack can achieve successful misclassification at different victim camera locations.  We show this for both day and night conditions, with our laser up to 25 meters away from the target sign, and using only 26 mW of laser power, in~\S\ref{sec:ir_cap} and~\S\ref{sec:evaluation}.
In contrast, LED light beams diverge in flight, attenuating over long distances.  This makes them unsuitable for long-distance, confined pattern projection attacks unless a high powered beam is aimed directly at a vehicle's camera.  As an example,  ICSL~\cite{wang2021can}, an IR LED-based attack, requires its LED light to operate at 30 Watts at 12 meters and must be aimed directly at the victim vehicle's camera (which is likely in motion) in order to successfully create fake objects.

\textit{(2) Continuous Tracking.}
A major challenge of light projection-based attacks on cameras, such as GhostImage~\cite{259721}, Rolling Colors~\cite{yan2022rolling}, and similar laser spoofing attacks on LiDARs (PLA-LiDAR~\cite{jin2023pla}, PRA~\cite{cao2023you}, Adv-LiDAR~\cite{cao2019adversarial}, and Illusion and Dazzle~\cite{illuanddazz}), 
 is the requirement to accurately and continuously aim at the victim sensor. This step is essential to ensure timing synchronization and accurate projection placement, which are needed to generate the correct adversarial pattern. 
Nevertheless, accurately tracking the sensor location on a CAV at any given moment proves challenging due to the dynamic nature of the vehicle's motion and external disturbances like vibrations, rendering the execution of such attacks difficult in practical scenarios.

To overcome this challenge, the use of reflection instead of direct injection has become a viable strategy in recent proof-of-concept attacks. For instance, AdvLB~\cite{duan2021adversarial} and AdvSL~\cite{LI2023103345} project visible light spots onto objects to fool object detectors and traffic sign recognition models. 
However, such attacks use human-detectable visible light and have not shown effectiveness in real-world moving driving scenarios.
In contrast, we realize that optimized invisible laser reflections are capable of inducing stable and prolonged misclassification in moving CAVs by targeting single-stage and two-stage traffic sign recognition architectures without the need for tracking and accurate projection.
We demonstrate the effective application of our technique to moving vehicle scenarios in~\S\ref{sec:eval_outdoor}.

\textit{(3) Ambient Light Variations.}
Another challenge of light-based attacks is their effectiveness under different lighting conditions. Previous work that used projected lights~\cite{nassi2020phantom, lovisotto2021slap} and shadows~\cite{zhong2022shadows} only succeeded under specific lighting conditions, such as at night. Success depends in part on the light source used. For example, non-coherent LED sources, as well as diffuse light from projectors, undergo scattering, causing it to spread out. When there is strong ambient light, the increased total luminance reduces the contrast between the projected pattern and its background. This can significantly reduce the attack success rate, rendering the attack impractical in bright environments. We show how ILR can succeed under diverse lighting conditions in~\S\ref{sec:attack_robust}.

\noindent\textbf{Defense Strategies.}
Wang et al.~\cite{wang2021can} propose a defense against the ICSL attack. Their defense strategy distinguishes active street lights from IR light sources. Unlike active street lights, IR light sources do not reflect off of roadways. If the source lacks a reflection, it is not an active street light. They also use differences in reflection colors for real street lights versus the non-reflected sources used for attacks in order to distinguish between active and IR sources. Our attack can not be detected or mitigated using their solution since, unlike active street lights, street signs lack active illumination.

The ILR attack is a type of perception attack that changes how a small portion of a target traffic sign is perceived, as shown in Fig.~\ref{fig:demo}.  Thus, it can be considered an adversarial patch attack~\cite{brown2017advpatch, eykholt2018physical, chen2018shapeshifter, zhao2018seeing}.
Defenses against adversarial patch attacks should apply in theory. So far, there are two types of defenses against adversarial patch attacks: (i) empirical defenses, such as the detection of anomalous patterns in attack patches~\cite{hayes2018visible, gowal2019scalable, madry2017towards, zhao2018seeing, yu2021defending, zhong2022shadows}, and (ii) certified defenses with theoretical guarantees~\cite{Chiang2020Certified, xiang2022patchcleanser, xiang2021patchguard, levine2020randomized}. Since the former is vulnerable to adaptive attacks~\cite{Chiang2020Certified}, we focus on certified defenses, especially PatchCleanser~\cite{xiang2022patchcleanser}, which is the current state-of-the-art. We evaluate our ILR attack against this defense and propose potential alternative defense strategies in~\S\ref{sec:attack_effectiveness}.

\nsection{Threat model and attacker capabilities} 
\label{sec:threat_model}

Fig.~\ref{fig:demo} shows an overview of the ILR attack. The attacker's goal is to cause the vision-based traffic sign recognition system of a victim CAV to incorrectly classify a target traffic sign, such as a stop sign, as a different type of sign, such as a yield sign.  We focus on untargeted attack scenarios. Sign misclassification can cause the CAV to behave dangerously, such as by braking unexpectedly or not stopping at an intersection. To do this, the adversary uses an IR laser to project an infrared light pattern onto a target sign with a specific size and position relative to that sign.  While invisible to humans, the IR laser's reflected pattern is visible to the CAV's camera sensor.  This results in the perception system's sign misclassification.

\noindent\textbf{Prior Knowledge and Assumptions.}
We assume that the attacker can obtain specifications for the victim camera, such as the presence of the IR filter, by using public information such as datasheets and teardown reports~\cite{auditeardown, bmwteardown}. This is similar to the assumptions made by prior attacks on CAV cameras~\cite{yan2022rolling, wang2021can}. Note that we assume the camera's internal settings, such as exposure time, are unknown to the attacker.

We also assume that the attacker has a basic understanding of IR light and optics in order to control the location of the projected IR pattern on the traffic sign surface as in previous work~\cite{cao2019adversarial, cao2023you}. The attack is remote and does not require any firmware access or information about the images captured by the camera in the victim CAV. However, we assume the attacker has access to, or can purchase, a similar camera, and can empirically study and infer the properties of the camera as in previous work~\cite{yan2022rolling, kohler2021they}. We assume that the attacker has knowledge of the traffic sign recognition model used by the victim CAV and has black-box access to it as an oracle (for example, by reverse-engineering the vehicle communication~\cite{jing2021tencent}). Specifically, the attacker can learn the recognition results, including the confidence scores, of similar models but cannot directly access the victim CAV's model or its internal parameters (e.g., model weights). 

\noindent\textbf{Attack Scenarios.}
The attacker selects a target traffic sign and places an IR laser emitter in line of sight, up to 25 meters away, based on our evaluation setup, from the traffic sign along the roadside where the victim CAV is likely to pass by. The attacker can find a suitable location in advance, and the attack device can be secretly installed to reduce the possibility of being detected by others, as was done in prior work~\cite{lovisotto2021slap, nassi2020phantom}. Note that the suitable location of the emitter is adjustable by the attacker using appropriate lenses and supports.

\begin{figure}[t!]
  \begin{minipage}[t!]{.99\linewidth}
    \includegraphics[width=\linewidth]{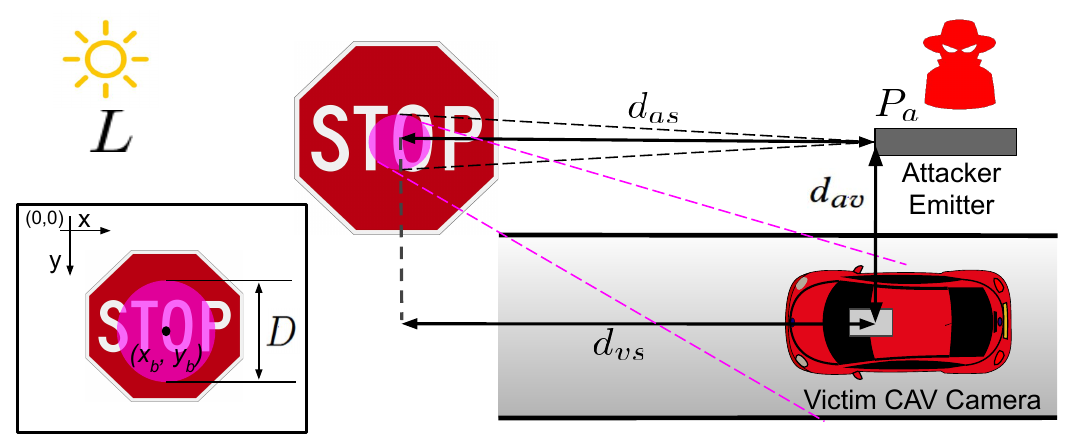}
    \caption{Overview of parameters of ILR attack
    }
    \label{fig:threat_model}
  \end{minipage}
\end{figure}

\nsubsection{Attack Modeling} \label{sec:attack overview}
We formulate the ILR attack as in Fig.~\ref{fig:threat_model}, where the distance of the victim AV camera from the traffic sign, $d_{vs}$, and the distance of the victim AV camera from the attacker IR emitter $d_{av}$ change dynamically as the victim CAV moves. We model the ILR attack with the parameters listed in Table~\ref{tbl:variables}.

The \textit{attack parameters} represent the factors controlled by the attacker and include (1) the distance of the attacker's laser from the traffic sign, $d_{as}$; (2) the laser beam power (in mW), $P_a$; (3) the diameter of the projected IR pattern, \textit{D}; and (4) the location of the center of the IR pattern in the traffic sign surface coordinates, $(x_b, y_b)$. The attacker can optimize various combinations of parameters to maximize the attack's effectiveness. Throughout this work, we consider a circular IR pattern with a diameter $D$ to evaluate our ILR attack. This is because it is the easiest pattern to create for a non-sophisticated attacker by positioning a lens or an iris in front of the laser emitter. Thus, we define the \textit{size} of the projected pattern in terms of the diameter value $D$.
More elaborate shapes and patterns can be achieved with more specialized equipment.

Finally, \emph{scenario parameters} represent environmental factors not controllable by the attacker, such as the ambient light intensity $L$, the longitudinal distance between the victim camera and the traffic sign $d_{vs}$, and the lateral distance between the victim camera and the attacker setup $d_{av}$, as shown in Fig.~\ref{fig:threat_model}. 
Note that we define only the minimum set of required parameters in our attack formulation necessary for the attacker to pursue a successful attack, as demonstrated in~\S\ref{sec:evaluation}.

\begin{figure}
\begin{center}
    \includegraphics[width=1\linewidth]{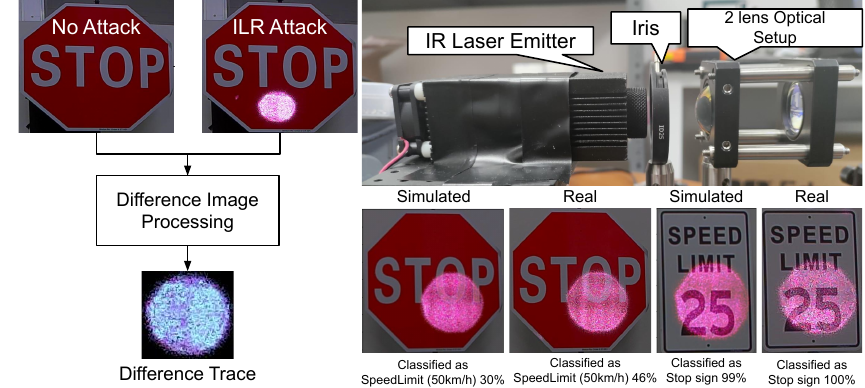}
\end{center}
\vspace{-0.1in}
\caption{Overview of Image Difference-based IR Trace Modeling (Left). The IR Laser module has a two-lens optical setup (right-top). A comparison of the simulated IR pattern with our modeling and the corresponding real-world IR pattern.}
\label{fig:simulation_demo}
\end{figure}

In our analysis, we consider the CAV camera output to be a stream of still images, and we call the set of still image pixels altered by our ILR attack an \textit{attack trace} (see Fig.~\ref{fig:simulation_demo}).  Every change in the attacker parameters in Table~\ref{tbl:variables} independently affects the camera's output image, resulting in different attack traces.
Using these assumptions, we describe an attack model optimization that accounts for \textit{temporal image noise} (random fluctuations in victim camera's output) in terms of image pixel intensity (photon count) values~\cite{896233} that occur as stray photons hit different sensor photosites across consecutive image frames. We thus evaluate the attack's effects on CAV sign classification over multiple consecutive camera image frames. A detailed methodology description is provided  in~\S\ref{sec:methodology}.

\begin{table}[t!]
\vspace{0.2in}
\centering
\scriptsize
\setlength{\tabcolsep}{1.8pt}
\setlength{\aboverulesep}{0pt}
\setlength{\belowrulesep}{0pt}
\renewcommand{\arraystretch}{1.1}
\caption{Definition of parameters} \label{tbl:variables}
\begin{minipage}[t]{.48\linewidth}
\centering
\begin{tabular}{cc} \toprule
Attack      & Parameter \\
Parameters     & Description \\ \midrule
$d_{as}$     & Distance: attacker $\leftrightarrow$ sign   \\
$D$ & Diameter of IR pattern  \\
 $P_{a}$   & Laser power   \\
$(x_b, y_b)$  & IR Pattern center coordinates   \\ \bottomrule
\end{tabular}
\end{minipage}
\hfill
\begin{minipage}[t]{.48\linewidth}
\centering
\begin{tabular}{cc} \toprule
Scenario   & Parameter \\
Parameters     & Description \\ \midrule
$d_{vs}$ &  Distance: victim $\leftrightarrow$ sign\\
$d_{av}$ & Distance: attacker $\leftrightarrow$ victim\\
$L$ & Intensity of ambient light \\
& \\ \bottomrule
\end{tabular}
\end{minipage}
\vspace{-0.2in}
\end{table}

\nsubsection{Physics of IR Laser Reflections}
\label{sec:ir_feasibility}
To define the attacker's capability to conduct the attack, it is necessary first to understand the impact of the reflected, projected IR pattern on the output images of the CAV camera. 
As described in~\S\ref{sec:attack_tsd}, laser light is a coherent source where all the light waves are in phase. In contrast with diffuse light, laser light preserves some properties of the original beam when reflected. Thus, when a laser beam strikes an ideal reflective surface, the reflected beam will be similarly directional and focused, preserving all the properties of the original beam.

\begin{figure}[t!]
\begin{center}
    \includegraphics[width=1\linewidth]{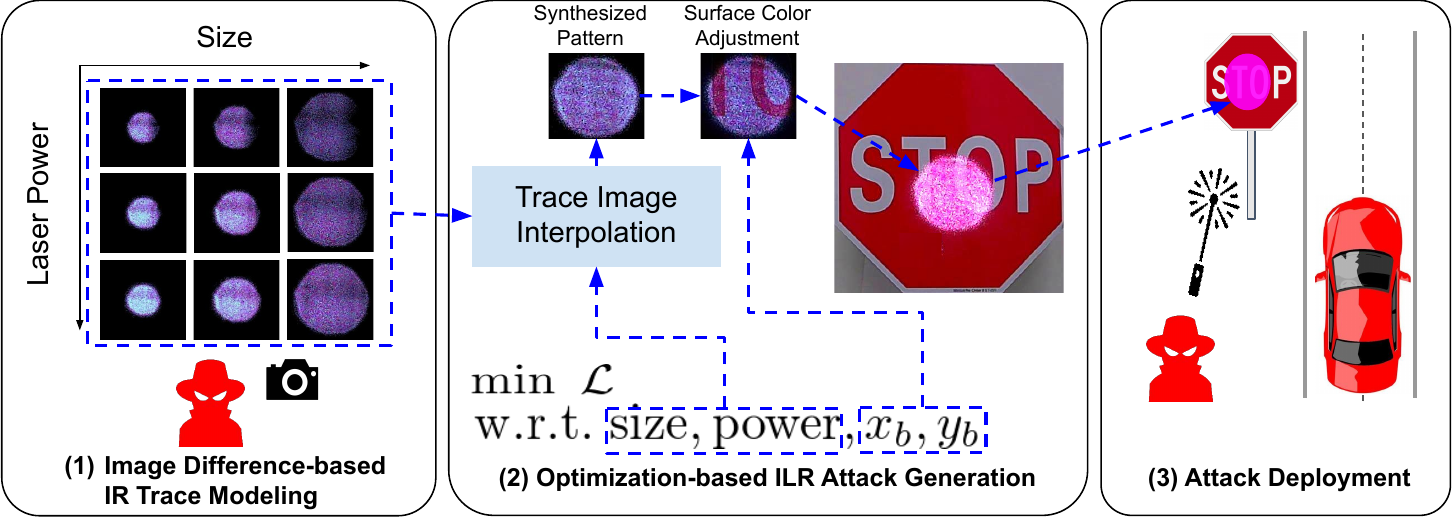}
\end{center}
\vspace{-0.1in}
\caption{Overview of ILR attack generation. (1) The attacker first collects IR traces on the targeted traffic sign for different laser powers and sizes, and (2) optimizes the attack w.r.t size, power, and location of the projected IR pattern. We apply a color adjustment based on the base color of the traffic sign. (3) The attack is deployed and verified in real-world scenarios.}
\label{fig:ilr_attack_overview}
\end{figure}

Traffic signs, such as the ones in Fig.~\ref{fig:demo}, generally use high-quality corrosion-resistant aluminum alloy sheets to meet  Manual on Uniform Traffic Control Devices (MUTCD) standards~\cite{fhwa}. When a laser beam illuminates a rough surface that is not perfectly reflective, such as a traffic sign, the laser beam is scattered in all directions by surface irregularities. These scattered light waves interfere with each other and produce a \textit{speckle pattern} visible in the CAV camera from different viewing angles. Since the scattered waves are still coherent, they preserve the same directionality and circular shape as the incident laser beam.  A small portion of the light is diffused, adding noise to the captured image while also attenuating the light, as shown in Fig.~\ref{fig:ilr_attack_overview}.  The ratio of coherent to incoherent light depends on surface roughness, laser light wavelength, and camera locations, all of which affect the visibility and complexity of the speckle pattern~\cite{tan2021specularity}.

\nsubsection{ILR Attack Capability Study} \label{sec:ir_cap}

In this section, we investigate the ability of an attacker to perform an ILR attack. 
Our study aims to define  (1) the relationship between the emitted IR power and the resulting range of pixel intensity variations in the captured images, (2) the correlation between the size of the projected IR pattern and image pixel intensity variation in the resulting speckle, and (3) the maximum achievable distance from which the attacker can make a successful attack. 
Note that the experiments in this section are conducted in a controlled, closed, indoor environment, except for the maximum distance experiment. We use real-world aluminum stop, and 25 mph speed limit, signs as targets.  We also use a Leopard camera with an OnSemi AR032ZWDR image sensor as the victim camera~\cite{leopard_camera} (referred to as OnSemi in the rest of the paper). OnSemi's camera is an automotive camera used by Baidu Apollo~\cite{apollo_hardware}. 

\noindent\textbf{Attacker Setup.}
\label{sec:attack_setup} The attacker setup, used in all experiments in this work, consists of an IR laser emitter that projects an IR pattern with a controllable size onto the target traffic sign's surface. We use a 780 nm IR laser module from CivilLaser~\cite{civil_IRmod} with a maximum output power of 1 W. It projects a collimated beam with a 0.7 cm diameter and 0.75$^{\circ}$ divergence angle. The attacker controls the power $P_a$ of the laser by changing the input current to the laser module.
Laser modules generally consist of a stack of multiple edge-emitting laser diodes. These diodes have different parallel and perpendicular divergence angles, resulting in an elliptical beam~\cite{4711324}. 
Thus, we place an iris in front of the laser module to create a circular IR pattern from the original elliptical beam.  Finally, the divergence angle of the projected laser beam is regulated by adjusting the distance between a two-lens optical setup as shown in Fig.~\ref{fig:simulation_demo}. This design allows the attacker to control the projected circular pattern diameter $D$. Details on laser safety are described in~\S\ref{sec:discussion} and Appendix~\ref{appendix:safety}.

\begin{figure}[t!]
\begin{center}
    \includegraphics[width=0.8\linewidth]{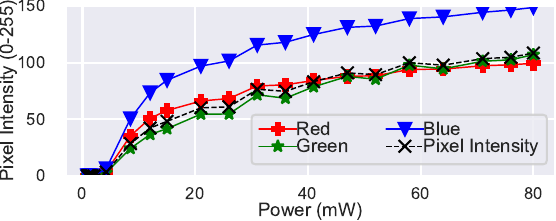}
\end{center}
\begin{center}
    \includegraphics[width=0.8\linewidth]{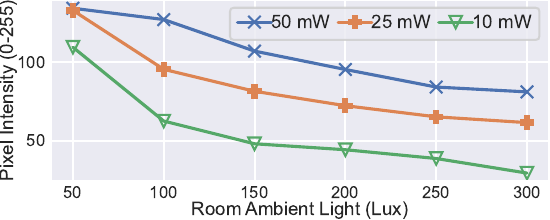}
\end{center}
\vspace{-0.1in}
\caption{8-bit RGB pixel intensity and overall pixel intensity variation of the attack traces for $D=15$ cm IR pattern at increasing power (top). The impact of artificial ambient light on pixel intensity offset (bottom).}
\label{fig:capability}
\end{figure}

\noindent\textbf{Laser Power vs Pixel Intensity.} In a controlled, indoor scenario, we set $d_{as} = 3$ m, the maximum distance achievable indoors. The room is illuminated by artificial ambient light, $L$, at 100 Lux. We then position the victim OnSemi camera such that $d_{av}$ is 0.3 m and $d_{vs}$ is 3 m. We set the circular IR pattern's diameter $D$ to 15 cm and, starting from 0 mW, we increase the power up to 80 mW and measure the average difference in RGB pixel values created by the speckle as illustrated in Fig.~\ref{fig:capability}.
We observe that the minimum laser power required for the attacker to alter the image's pixel values and create a speckle is 2.4 mW -- less than the power emitted by a laser pointer. This can be achieved by operating our laser module at 0.25\% of its capability. We further notice that the 8-bit intensity variation created for blue pixels is larger than for red and green pixels (by at least 30) for laser powers greater than 20 mW. The red and green channel intensity variations follow a similar trend, as shown in Fig.~\ref{fig:capability} (top). 

Similar to the laser power variation, the room's ambient light impacts pixel intensities by varying the victim camera's auto-exposure-controlled light sensitivity. Note that we treat the victim camera as a black box. The bottom graph in Fig.~\ref{fig:capability} shows that the average 8-bit pixel intensity variation for the attack traces decreases with increasing room ambient light, $I$, at a logarithmic rate of $-40.1 \cdot \log(I)$ with a measured offset based on the transmitted laser beam power.

\begin{figure}[t!]
\begin{center}
    \includegraphics[width=0.8\linewidth]{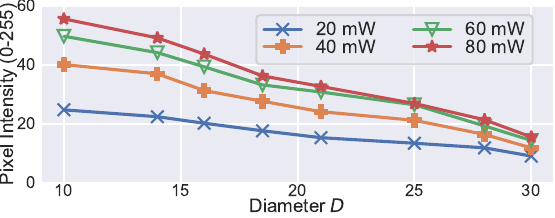}
\end{center}
\vspace{-0.1in}
\caption{The offset in 8-bit pixel intensity values at increasing IR pattern size $D$ at different laser powers $P_a$.}
\label{fig:size_capbility}
\end{figure}

\noindent\textbf{IR Pattern Size vs Pixel Intensity.}
The attacker can use the two-lens optical setup to control the size of the projected circular IR  pattern. For a given distance $d_{as}$, we observe that the attacker can achieve a circular diameter $D$ between 3.5 cm and 30 cm based on our hardware setup. The details about laser power attenuation with respect to beam size increase are provided in Appendix~\S\ref{appendix:laser_power_attenuation}.
We then characterize the pixel intensity variation for laser power increasing from 20 mW to 80 mW.  We observe that the intensity offset decreases linearly at increasing sizes of IR patterns, as shown in Fig.~\ref{fig:size_capbility}.

\noindent\textbf{Attenuation at Increased Lateral Distance.}
We verify the speckle intensity attenuation in the captured images at increasing lateral distance from the emitter ($d_{av}$). 
We place the emitter at 3 m ($d_{as}$) away from a stop sign and then measure the variation of pixel intensity values in the RGB channels of the captured attack trace images.
We observe pixel intensity drops of up to 18\% when the victim camera moves from 0 to 1.5 m (approximately the distance from the center of a roadway lane).  Since this attenuation is negligible compared to the attenuation due to IR pattern size and emitter distance from the target sign, we only consider those factors in our attack optimization design described in~\S\ref{sec:methodology}.

\nsection{ILR Optimized Attack Methodology}
\label{sec:methodology}

Based on the attacker capabilities described in~\S\ref{sec:ir_cap}, we design an optimization framework to automatically generate effective ILR attacks in terms of the optimal IR pattern location, the minimum circular pattern diameter, and the minimum laser power required by the attacker to achieve misclassification. Fig.~\ref{fig:ilr_attack_overview} shows an  ILR attack generation overview. To obtain optimized attacks, the framework performs: (1) image difference-based IR trace modeling (\S\ref{sec:diff_modeling} and \S\ref{sec:trace_interpolation}), (2) optimization-based ILR attack generation (\S\ref{sec:attack_opt}), and (3)  attack deployment on attack scenarios (\S\ref{sec:evaluation}).

\nsubsection{Image Difference-based IR Trace Modeling} \label{sec:diff_modeling}
To optimize ILR attack effectiveness, we first synthesize the attack traces captured by the victim camera. As described in~\S\ref{sec:threat_model}, the IR pattern projection generates a speckle (attack trace) in the output images.
Accurately synthesizing attack traces is challenging since they result from multiple, randomly phased, coherent waves. Thus, we model the phenomena by collecting and applying image differencing to the attack traces to extract RGB intensity offsets while varying attacker parameters, such as emitted power and circular IR pattern size, as shown in Fig.~\ref{fig:simulation_demo}. More details are in Appendix ~\ref{appendix:empirical}.

\subsection{Trace Image Interpolation} \label{sec:trace_interpolation}

\begin{figure}[t!]
\begin{center}
    \includegraphics[width=0.95\linewidth]{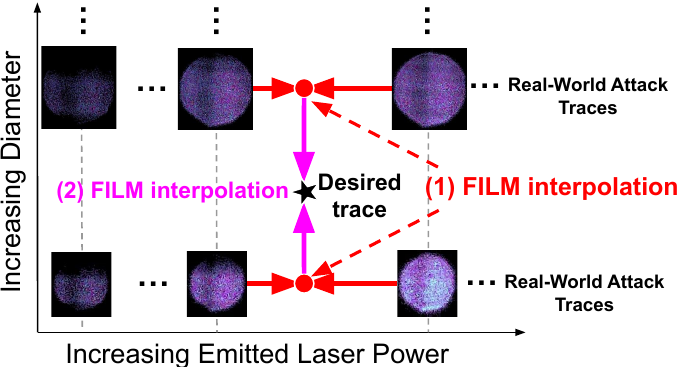}
\end{center}
\vspace{-0.1in}
\caption{Overview of the DNN-based interpolation using FILM~\cite{reda2022film}. The method interpolates two traces at increasing emitted power and IR pattern diameters $D$.}
\label{fig:interpolation_methods}
\end{figure}

Collecting all possible real-world traces for all laser powers and IR pattern sizes is infeasible. 
Naive interpolation with averaging does not work for our attack, as averaging cancels out the speckle patterns. For this reason, we design a method to derive attack traces by interpolating real-world traces. To preserve the local spatial information while interpolating, we adopt a recent DNN-based frame interpolation algorithm, FILM~\cite{reda2022film}. FILM generates slow-motion videos from very similar photos. As shown in Fig.~\ref{fig:interpolation_methods}, we build the interpolation process for increasing laser powers and trace diameters and obtain intermediate attack traces as video frames. We note that this methodology can be applied to different traffic signs by adjusting the trace color or by collecting the appropriate real-world traces. We use the official pre-trained model in~\cite{reda2022film}.

\nsubsection{Optimization-based ILR Attack Generation} \label{sec:attack_opt}

Finally, we design a black-box optimization formulation to optimize the image difference-based IR trace modeling  (\S\ref{sec:diff_modeling}) and trace image interpolation (\S\ref{sec:trace_interpolation}). This technique allows the simulation of an attack-influenced image with arbitrary trace position $(x_b, y_b)$, laser power $P_{a}$, and trace diameter $D$, to find the optimal configuration. 
For other parameters listed in Table~\ref{tbl:variables}, we do not directly optimize the parameters, but consider them in the expectation over transformation (EoT) technique~\cite{athalye2018synthesizing, brown2017advpatch} to be robust against changes to them. The attack formulation can be written as follows:

\vspace{-0.2in}
\begin{align}
    \min \ \ \ 
    & \mathbb{E}_{X \sim \operatorname{EoT}(X_{ILR})} \left [ \mathcal{L}(X, \theta) \right ] \nonumber\\
    {\rm s.t.} \ \ \ 
    & X_{\rm ILR} = \operatorname{TraceModeling}(X_{\rm base}, T, x_b, y_b) \label{math:attack_opt},  \\ \nonumber
    & T = \operatorname{Interp}(D, P_a) 
\end{align}

where the diameter of trace $D$, laser power $P_a$, and the position of attack trace $(x_b, y_b)$ are the decision variables and $\theta$ is the targeted DNN model's parameter set. Interp$(\cdot)$ is a function of the trace image interpolation.  $\operatorname{TraceModeling}(\cdot)$ is a function of the image difference-based IR trace modeling used to get a simulated attack-influenced image, $X_{\rm ILR}$.  $X_{\rm base}$ is a benign (base) image containing the target traffic sign. 
In $\operatorname{EoT}(X_{\rm ILR})$, we sample images with the EoT technique from the simulated image $X_{\rm ILR}$. 
In the EoT, we add Gaussian noises, change color brightness, and apply rotation and shear. $\mathcal{L}$ is a loss function. In this study, we simply minimize the confidence value of the target class ---  our attack is an untargeted attack as discussed in~\S\ref{sec:threat_model}.
As the attack formulation Eq.~(\ref{math:attack_opt}) is not differentiable, we use a black-box optimization method to find effective $D, P_a$, and $(x_b, y_b)$. We adopt the Tree-structured Parzen Estimator algorithm~\cite{bergstra2011algorithms} in Optuna~\cite{optuna_2019}. We note that the optimization generally converges to the global minimum since the search space is small (4 variables).

\nsection{Evaluation} \label{sec:evaluation}
We evaluate the ILR attack for effectiveness, generality, robustness, and transferability in the real world. We also evaluate the effectiveness of ILR attacks in outdoor moving victim scenarios.

\nsubsection{Attack Effectiveness and Generality Evaluation} \label{sec:attack_effectiveness}

In this section, we evaluate our attack on real-world aluminum, stop, and 25 mph speed limit signs in a controlled, closed, indoor environment with the setup described in~\S\ref{sec:ir_cap}, as shown in Fig.~\ref{fig:demo}.
We place the victim camera and IR emitter at $d_{vs}=d_{as}=3$ m in front of the target sign. For collecting the traces for our optimization model, we increase the laser power $P_a$ from 2.4 to 80 mW and the diameter $D$ from 10 to 30 cm, based on our assumed attacker capabilities. The artificial ambient light, $L$ is set to 100 Lux. We use the OnSemi camera as a default for this evaluation if not mentioned otherwise. Table~\ref{tbl:cameras} lists all the four cameras tested in the generality study.

Note that we evaluate the physically realized ILR attack in the real world after applying our optimization methodology, not the digitally simulated IR patterns. We then compare our results against a baseline random attack in which an IR laser beam hits random portions of the target signs. Detailed setup of the random attack is in Appendix~\ref{appendix:random_attack}.

\begin{table}[htb]
\vspace{0.2in}
\centering
\footnotesize
\setlength{\tabcolsep}{1.8pt}
\setlength{\aboverulesep}{0pt}
\setlength{\belowrulesep}{0pt}
\renewcommand{\arraystretch}{1.1}
\caption{Benign Performance of the object detectors and classifiers for traffic sign recognition. 
The architecture of the CNN model is in Appendix~\ref{appendix:model_arch}.
YOLOv3 is evaluated in APb. Others are in mAP.}
\scalebox{0.895}{
\begin{tabular}{ccccc}\toprule
Object Detector (Training Dataset)     & mAP/APb &  &    Classifier (Training Dataset)         & Acc. \\ \cline{1-2} \cline{4-5} 
Faster-RCNN~\cite{ren2015faster} (ARTS~\cite{almutairy2019arts})      & 84.3     &  & CNN (ARTS~\cite{almutairy2019arts})  & 81\% \\
Faster-RCNN~\cite{ren2015faster} (Mapillary~\cite{ertler2020mapillary}) & 18.3     &  & CNN (LISA~\cite{mogelmose2012vision})  & 99\% \\
YOLOv3~\cite{redmon2018yolov3} (COCO~\cite{lin2014microsoft})           & 33.8     &  & CNN (GTSRB~\cite{Houben-IJCNN-2013}) & 98\% \\ \cline{1-2} \cline{4-5}  \toprule
\end{tabular}}
\label{tbl:benign_performance}
\end{table}

\nsubsubsection{Targeted Traffic Sign Recognition Models}
Table~\ref{tbl:benign_performance} lists the targeted object detectors and classifiers and their benign performances.
For the \textit{single-stage architectures}, we train an object detection model with the ARTS~\cite{almutairy2019arts} and Mapillary~\cite{ertler2020mapillary} datasets. As the Mapillary dataset includes worldwide traffic signs, we only use the stop and speed limit signs used in the United States. For the stop sign only, we evaluate YOLOv3~\cite{redmon2018yolov3} trained with the COCO dataset~\cite{lin2014microsoft}, which is a generic object detector but does not contain US-style speed limit signs. Thus, we evaluate different datasets for stop and speed limit signs. We use 0.3 as the object-detection threshold of confidence score, following conventional practice~\cite{mmdetection}.

For the \textit{two-stage architectures}, we manually crop the ROI area for each sign to focus on the analysis of the second-stage classification.
For the second-stage classifiers, we train classification models with three datasets, one trained on European traffic signs and the other on U.S. signs.
For the European traffic signs, we trained a CNN classification model on the GTSRB~\cite{Houben-IJCNN-2013} dataset. For the U.S. traffic signs, we trained CNN models on the LISA~\cite{mogelmose2012vision} and ARTS~\cite{almutairy2019arts} datasets. We use a CNN model architecture that is among the best performers on the GTSRB dataset~\cite{cnn_gtsrb}. Details are in Appendix~\ref{appendix:model_arch}.

\begin{table}[t!]
\centering
\footnotesize
\setlength{\tabcolsep}{3.5pt}
\setlength{\aboverulesep}{0pt}
\setlength{\belowrulesep}{0pt}
\renewcommand{\arraystretch}{1.1}
\caption{Target cameras considered in our evaluation.}
\begin{tabular}{ccccc}
\toprule
           & \textbf{Leopard}   & \textbf{Raspberry Pi}    & \textbf{LifeCam} & \textbf{Leopard}\\
           & \textbf{OnSemi}   & \textbf{HQ v1.1}    & \textbf{HD-3000} & \textbf{OmniVision}\\
           & \includegraphics[height=0.5in]{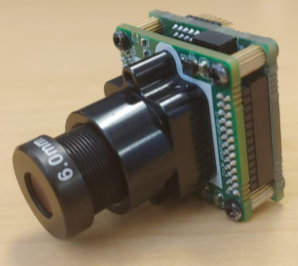}          &     \includegraphics[height=0.5in]{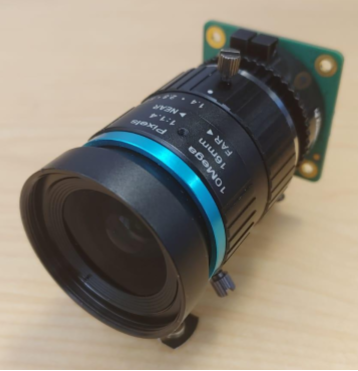}         &        \includegraphics[height=0.5in]{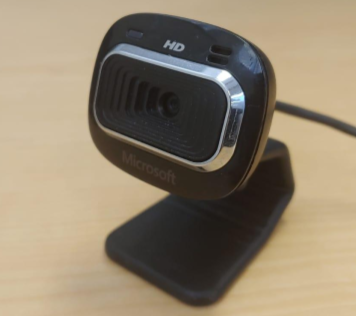}     &
           \includegraphics[height=0.5in]{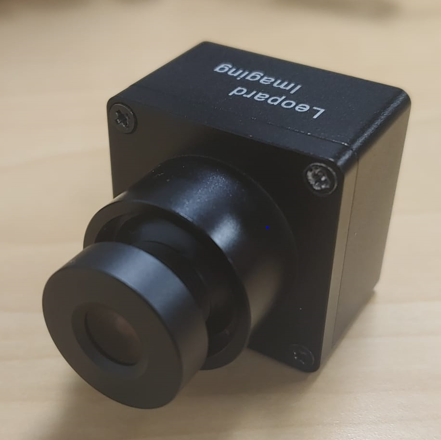} \\ \midrule
Sensor     & AR032ZWDR & IMX477       & N/A       & OV10635  \\
Usage &       CAV & General &  WebCam & CAV  \\
Resolution & 1920$\times$1080 & 4056$\times$3040  & 1280$\times$720    &  1280$\times$800   \\
Lens $f$     & 6 mm       & 16 mm         & N/A    & 6 mm           \\
Max FPS    & 30        & 90           & 30      & 30          \\
FOV        & H - 60$^{\circ}$     & 44.6$^{\circ}$ $\times$ 33.6$^{\circ}$ & D - 68.5$^{\circ}$ & D - 68.5$^{\circ}$          \\ \bottomrule
\end{tabular}
\label{tbl:cameras}
\end{table}

\nsubsubsection{Evaluation Metrics}
We design two evaluation metrics based on our threat model (\S\ref{sec:threat_model}) and attack design: the attack success rate (ASR) and the simulation consistency rate (SCR). ASR measures the percentage of cases in which a sign is misclassified or undetected, thus satisfying our attack goal, as discussed in~\S\ref{sec:trace_interpolation}. SCR is defined as the percentage of cases in which the classification caused by the ILR attack is consistent between physical and digital scenarios. We use SCR to evaluate the quality of attack trace modeling (\S\ref{sec:attack_opt}). We do not consider the SCR for the single-stage architecture since a successful attack typically just prevents the object detector from detecting the traffic signs. ASR is our primary metric. SCR is necessary to evaluate the validity of our attack design.
\nsubsubsection{Results}\label{sec:eval_result_effectivness}
Table~\ref{tbl:attack_effectivness} shows the effectiveness of the ILR attack against single-stage and two-stage traffic sign recognition systems. Our ILR attacks show significantly higher effectiveness than the random attack, with a 100\% success rate for all models. 
These results indicate that while the IR laser traces fool traffic sign recognition systems, effective attack optimization is needed to cause a significant impact on recognition. Table~\ref{tbl:diff_interp} shows the ASR and SCR of the ILR attack compared with ``w/o interp.'', which only explores the discrete power and size values of the collected IR patterns without interpolation, and ``Spline Interp.'', which is a baseline method that adapts a cubic spline interpolation. Details on this method are described in Appendix~\ref{appendix:spline}.
The DNN-based interpolation method has the highest ASR and SCR with 100\% ASR and 92.5\% SCR on average. This reveals that high-quality interpolation is essential to find effective ILR attacks.

\begin{figure}[t!]
\begin{center}
    \includegraphics[width=\linewidth]{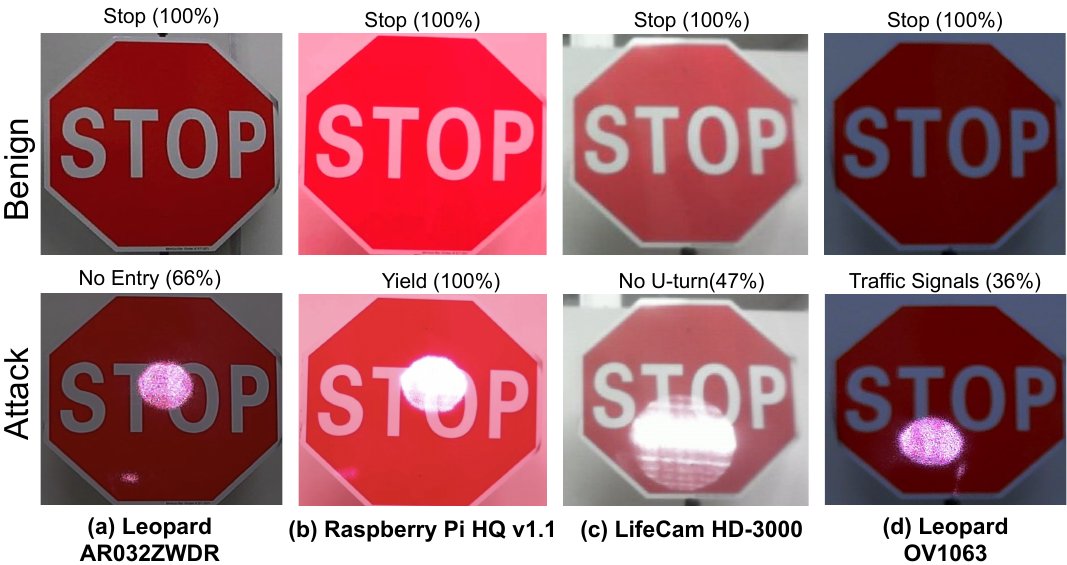}
\end{center}
\vspace{-0.05in}
\caption{Examples of captured images of the 4 IR-sensitive cameras in benign and during a successful attack. The detected colors differ as each camera has a different sensitivity to IR.}
\label{fig:camera_diff}
\vspace{0.1in}
\end{figure}

\noindent\textbf{Generality to Different Cameras.} \label{sec:eval_result_effectivness_camera} 
To further study the impact of the ILR attack on different cameras, we evaluate the attack's effectiveness using a Raspberry Pi HQ v1.1 camera~\cite{raspberrycam} with a Sony IMX477 image sensor~\cite{sonyIMX477}, a Microsoft LifeCam HD-3000 camera~\cite{microsoft_lifecam}, and another automotive camera with an OmniVision OV10635 image sensor~\cite{leopard2, OV10635} (referred to as OmniVision) with the IR filter removed. As shown in Fig.~\ref{fig:camera_diff}, the perceived IR pattern varies between different cameras, as they vary in their sensitivities to infrared wavelengths.
Table~\ref{tbl:other_cameras} lists the ASR and SCR of the ILR attack on the four tested cameras. The ILR attack is always successful, with an ASR of 100\%. On the other hand,  the speed limit sign SCR was 0\% for the Raspberry Pi HQ v1.1 and LifeCam cameras. We observed that IR trace color is camera sensor dependent.  Thus, the relationship between trace color, diameter, $D$, and power, $P_a$, is also sensor dependent.  This affects the accuracy of simulated IR traces.

\noindent\textbf{Maximum Achievable Distance.}
To evaluate the maximum achievable distance from the emitter to the target traffic sign, the experiment was conducted in both indoor and outdoor scenarios in a controlled environment. Using our setup, we verify the ASR against the speed limit sign using the LISA model~\cite{mogelmose2012vision} configured as described in \S\ref{sec:attack_effectiveness}. 
Our results show that with our minimal setup, ILR consistently succeeds (100\% ASR) up to 25 meters away from the target sign with a power of 26 mW. Long-range attacks are possible because of the laser beam properties described in \S\ref{sec:ir_feasibility}. Beyond 25 m, the speckle intensity loss and beam divergence prevent coherent pattern shape projection, dropping the ASR to zero. More sophisticated optics can be used to increase the attack range.

\noindent\textbf{Attack with Saturated IR Speckle.} The optimization methodology focuses on minimizing the laser power necessary for the attacker to achieve a successful attack. We can further evaluate ILR by considering an attacker whose goal is to achieve camera sensor saturation using the projected IR laser speckle. We consider an IR speckle to be saturated if $>50\%$ of the pixels in the pattern are saturated, meaning intensity = 255.  For this analysis, maintaining the same attack scenario as the maximum achievable distance evaluation, we optimize the trace only with respect to the trace diameter $D$ and the coordinate position of the center of the trace $(x_b, y_b)$. We increase the IR beam power to achieve saturation when the trace diameter is optimized.  Our results for the OnSemi camera exhibit a 100\% ASR in all evaluated two-stage models. We observe that the optimized trace diameter for stop sign attacks drops from 21.25 cm to 17.5 cm on average and from 31.5 cm to 27 cm for speed limit sign attacks under saturation conditions. This evaluation reveals how attackers can achieve high attack success independently of the tested model at the cost of increasing laser power.

\noindent\textbf{Generality to Different Laser Wavelengths} To evaluate the attack generality against different laser wavelengths, we conducted experiments with 830-nm and 980-nm laser modules (in addition to our 780 nm laser). For each of the laser modules, we collect the IR traces and optimize for the attack trace individually. We find that the ILR attack can achieve a 100\% ASR on stop and speed limit signs with both  tested laser modules. 37 and 17 mW laser powers were required for the 830 and 980 nm laser modules respectively to attack the stop sign. Similarly, 44 and 26 mW laser powers respectively were required in the case of speed limit signs. We observe that for higher frequency modules, a lower laser power is required to attack a stop sign. We hypothesize that this is because the IR traces created by high-frequency lasers tend to appear with a more blue-shifted (higher contrast) hue in the camera image, when compared with the stop sign's red surface color.

\begin{table}[t!]
\vspace{0.1in}
\centering
\footnotesize
\setlength{\tabcolsep}{3.7pt}
\setlength{\aboverulesep}{0pt}
\setlength{\belowrulesep}{0pt}
\renewcommand{\arraystretch}{1.2}
\caption{Attack effectiveness of single-stage and two-stage traffic sign recognition systems with ASR and SCR.}
\scalebox{0.895}[1]{
\begin{tabular}{cccccccc}
\toprule
 &                            &                         & \multicolumn{2}{c}{Random Attack} &  & \multicolumn{2}{c}{ILR Attack} \\ \cline{4-5} \cline{7-8} 
 &                            &                         & ASR              & SCR            &  & ASR            & SCR           \\ \hline\hline
\multirow{4}{*}{\rotatebox{90}{Stop Sign}}   & \multirow{2}{*}{Single-Stage} & Faster R-CNN (ARTS) & 0\% & N/A &  & \textbf{100\%} & N/A \\
 &                            & YOLOv3 (COCO)           & 0\%                & N/A            &  & \textbf{100\%}          & N/A           \\ \cline{2-8} 
 & \multirow{2}{*}{Two-Stage} & CNN (ARTS)              & 0\%              & N/A            &  & \textbf{100\%}          & \textbf{100\%}         \\
 &                            & CNN (GTSRB)              & 20\%             & N/A            &  & \textbf{100\%}          & 70\%          \\ \hline\hline
\multirow{4}{*}{\rotatebox{90}{Speed Limit\hspace{0.3em}}} & \multirow{2}{*}{Single-Stage} & Faster R-CNN (ARTS) & 60\% & N/A &  &  \textbf{100\%}     & N/A \\
 &                            & Faster R-CNN (Mapillary) & \textbf{100\%}                & N/A            &  &  \textbf{100\%}              & N/A           \\ \cline{2-8} 
 & \multirow{2}{*}{Two-Stage} & CNN (ARTS)              & 64\%             & N/A            &  & \textbf{100\%}          & \textbf{100\%}         \\
 &                            & CNN (LISA)             & 10\%             & N/A            &  & \textbf{100\%}          & \textbf{100\%}         \\ \bottomrule
\end{tabular}}
\label{tbl:attack_effectivness}
\vspace{-0.1in}
\end{table}

\begin{table}[t!]
\vspace{0.2in}
\centering
\footnotesize
\setlength{\tabcolsep}{3pt}
\setlength{\aboverulesep}{0pt}
\setlength{\belowrulesep}{0pt}
\renewcommand{\arraystretch}{1.2}
\caption{Evaluation of the interpolation method. ``w/o interp.'' only optimizes the attack with discrete laser powers and sizes without interpolation. ``Spline Interp.'' is a baseline spline-based method detailed in Appendix~\ref{appendix:spline}.
}
\scalebox{0.92}[1]{
\begin{tabular}{cccclcclcclcc}
\toprule
 &  & \multicolumn{2}{c}{DNN-based Interp.} &  & \multicolumn{2}{c}{w/o Interp.} & & \multicolumn{2}{c}{Spline Interp.} \\ \cline{3-4} \cline{6-7} \cline{9-10} 
                             &             & ASR   & SCR   &   & ASR   & SCR  & & ASR & SCR \\ \hline\hline
\multirow{2}{*}{\rotatebox{0}{\begin{tabular}[c]{@{}c@{}}Stop \\ Sign\end{tabular}}}   & CNN (ARTS)  & \textbf{100\%} & \textbf{100\%} &  & 20\% & 20\%  &  & \textbf{100\%} & \textbf{100\%}  \\ \cline{2-10} 
                             & CNN (GTSRB) & \textbf{100\%} & 70\%  &  & 90\% & 90\% &  & 80\% & 80\% &  \\ \hline\hline
\multirow{2}{*}{\rotatebox{0}{\begin{tabular}[c]{@{}c@{}}Speed\\Limit\end{tabular}}} & CNN (ARTS)  & \textbf{100\%} & \textbf{100\%} &   & \textbf{100\%} & \textbf{100\%} &  & \textbf{100\%} & 0\% &  \\ \cline{2-10} 
                             & CNN (LISA)  & \textbf{100\%} & \textbf{100\%} &  & \textbf{100\%} & \textbf{100\%}  &  & \textbf{100\%} & \textbf{100\%}\\ \bottomrule
\end{tabular}}
\label{tbl:diff_interp}
\end{table}

\begin{table}[t!]
\vspace{0.2in}
\centering
\footnotesize
\setlength{\tabcolsep}{3pt}
\setlength{\aboverulesep}{0pt}
\setlength{\belowrulesep}{0pt}
\renewcommand{\arraystretch}{1.2}
\caption{Attack effectiveness on the 4 different cameras.
}
\scalebox{0.75}[1]{
\begin{tabular}{ccccccccccccc}
\toprule
 &  & \multicolumn{2}{c}{OnSemi} &  & \multicolumn{2}{c}{Raspberry Pi HQ} &  & \multicolumn{2}{c}{LifeCam HD-3000 } & & \multicolumn{2}{c}{OmniVision}\\ \cline{3-4} \cline{6-7} \cline{9-10} \cline{12-13}
                             &             & ASR   & SCR   &  & ASR   & SCR   &  & ASR   & SCR  & & ASR & SCR \\ \hline\hline
\multirow{2}{*}{\rotatebox{0}{\begin{tabular}[c]{@{}c@{}}Stop \\ Sign\end{tabular}}}   & CNN (ARTS)  & \textbf{100\%} & \textbf{100\%} &  & \textbf{100\%} & \textbf{100\%} &  & \textbf{100\%} & \textbf{100\%} & & \textbf{100\%} & 20\% \\ \cline{2-13} 
                             & CNN (GTSRB) & \textbf{100\%} & 70\%  &  & \textbf{100\%}  & \textbf{100\%}  &  & \textbf{100\%} & \textbf{100\%} & & 90\% & \textbf{90\%}\\ \hline\hline
\multirow{2}{*}{\rotatebox{0}{\begin{tabular}[c]{@{}c@{}}Speed\\Limit\end{tabular}}} & CNN (ARTS)  & \textbf{100\%} & \textbf{100\%} &  & \textbf{100\%} & 0\%   &  & \textbf{100\%} & \textbf{100\%} & & \textbf{100\%} & \textbf{100\%}\\ \cline{2-13} 
                             & CNN (LISA)  & \textbf{100\%} & \textbf{100\%} &  & \textbf{100\%} & 0\% &  & \textbf{100\%} & 0\% & & \textbf{100\%} & \textbf{100\%}\\ \bottomrule
\end{tabular}}
\label{tbl:other_cameras}
\end{table}

\nsubsection{Attack Robustness Evaluation} \label{sec:attack_robust}

We evaluate the robustness of the ILR attack under varied ambient lighting conditions and camera positions using the same indoor setting as detailed in~\S\ref{sec:attack_effectiveness}.

\begin{figure}[t!]
\begin{center}
    \includegraphics[width=\linewidth]{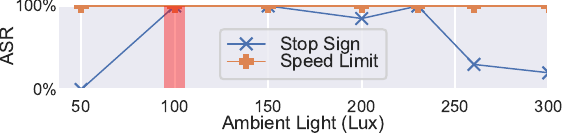}
\end{center}
\caption{Attack success rates under different ambient lights. The attack is generated under 100 Lux at the red bar.}
\label{fig:robust_ambient}
\end{figure}

\noindent\textbf{Robustness to Different Ambient Lightings.} \label{sec:attack_robust_light}
Fig.~\ref{fig:robust_ambient} shows the ASR of the ILR attack against second-stage classifiers with increasing ambient light levels. The attack is generated at 100 Lux and evaluated for 7 other artificial light levels, ranging from 50 to 300 Lux. For the stop and speed limit signs, we use the CNN classifier models trained with the GTSRB and LISA datasets, respectively.
As shown, the ILR attack against the stop sign shows high robustness between 100 and 230 Lux, but its ASR drops significantly above 230. In contrast, the attack against the speed limit sign shows high robustness, with 100\% ASR for all light settings. We believe the difference in performance is due to differences in contrast between the traffic sign surface colors and the laser speckle. On a white sign, the speckle has a higher contrast than on a red sign. The speckle color is dependent upon the laser wavelength and camera sensor used, as described in~\S\ref{sec:ir_feasibility}.

\noindent\textbf{Robustness to Different Object Detectors in Single-Stage Architecture.} \label{sec:attack_robust_pos_first}
Table~\ref{tbl:robust_camera_asr_first} lists the ASR for single-stage architecture object detectors at increasing distances $d_{vs}$ between the camera and the traffic sign. The attack is generated for all the models at a fixed distance ($d_{vs}$ = 6 m) and evaluated for the others. As shown, the ILR attack reaches high attack effectiveness for the speed limit sign with 100\% ASR at all tested distances and models. For the stop sign, the ILR attack is effective against Faster R-CNN trained on the ARTS dataset but not always effective against YOLOv3 and YOLOv5.
We believe these variations are due to the architectural differences in object detectors. The Faster R-CNN model, a two-shot object detector, finds region proposals and classifies those regions. It thus has a high ASR similar to the second-stage classification model as discussed in~\S\ref{sec:attack_effectiveness}. 
YOLOv3 and YOLOv5, single-shot object detectors, perform the two steps simultaneously. This strategy may improve robustness as it can take into account global features from the region proposal.
These results indicate that single-stage traffic sign recognition with a single-shot object detector can provide effective mitigation against ILR attacks. However, we note that the current object detectors are still not able to handle several types of different traffic signs, as discussed in~\S\ref{sec:overview_tsd}.

\begin{table}[t!]
\vspace{0.2in}
\centering
\footnotesize
\setlength{\tabcolsep}{7.5pt}
\setlength{\aboverulesep}{0pt}
\setlength{\belowrulesep}{0pt}
\renewcommand{\arraystretch}{1.1}
\caption{ASR for single-stage architecture under 4 different distances $d_{vs}$ between the camera and the sign.}
\scalebox{0.845}[1]{
\begin{tabular}{cccccc}
\toprule
Target Sign &       Detection Model                  & 4 m    & 5 m    & 6 m    & 7 m    \\ \midrule
\multirow{3}{*}{\begin{tabular}[c]{@{}c@{}}Stop\\ Sign\end{tabular}}   & Faster R-CNN (ARTS) & \textbf{100\%} & \textbf{100\%} & \textbf{100\%} & \textbf{100\%} \\
& YOLOv3 (COCO)           & 0\%   & 0\%   & \textbf{100\%} & 0\%   \\ 
& YOLOv5 (COCO)           & 10\%   & 90\%   & \textbf{100\%} & \textbf{100\%}   \\ 
\midrule
\multirow{3}{*}{\begin{tabular}[c]{@{}c@{}}Speed\\ Limit\end{tabular}} & Faster R-CNN (ARTS) & \textbf{100\%} & \textbf{100\%} & \textbf{100\%} & \textbf{100\%} \\
& Faster R-CNN (Mapillary) & \textbf{100\%} & \textbf{100\%} & \textbf{100\%} & \textbf{100\%} \\
& YOLOv5 (ARTS) & \textbf{100\%} & \textbf{100\%} & \textbf{100\%} & \textbf{100\%} \\ \bottomrule
\end{tabular}}
\label{tbl:robust_camera_asr_first}
\vspace{-0.1in}
\end{table}

\begin{figure}[t!]
\begin{center}
    \includegraphics[width=\linewidth]{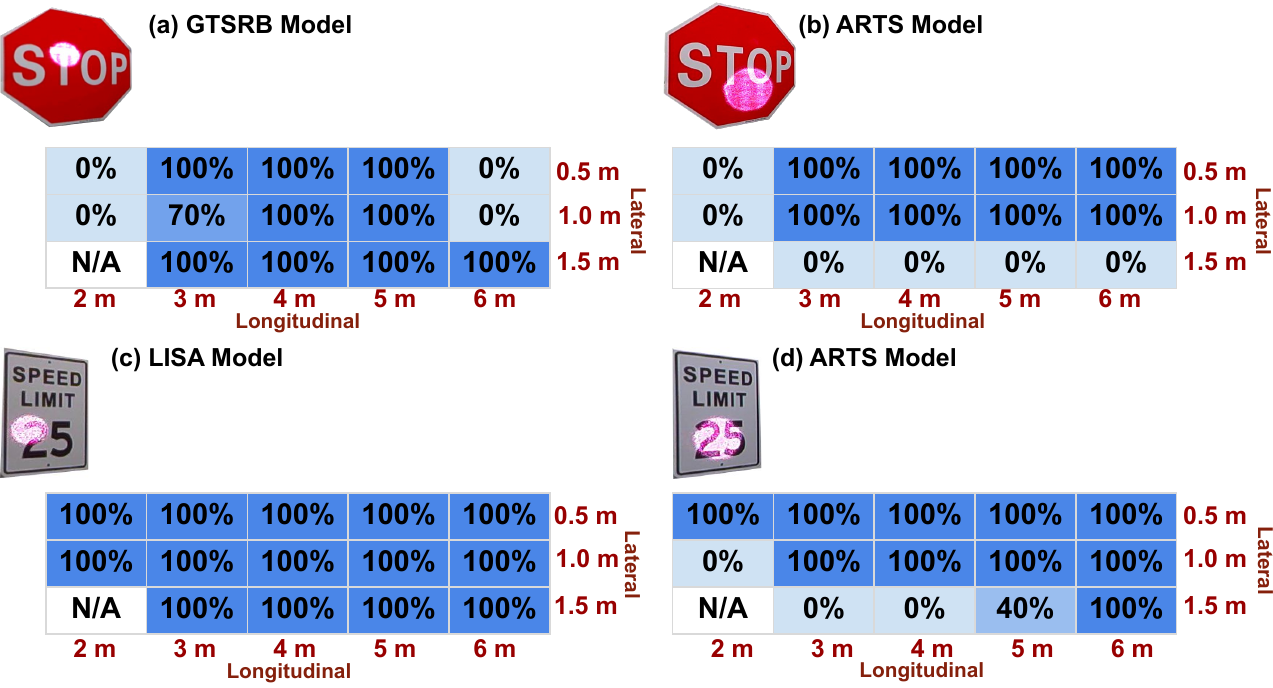}
\end{center}
\caption{ASR for two-stage architecture model with 14 different camera positions. N/A: the traffic sign is not visible.}
\label{fig:robust_camera_asr}
\end{figure}

\begin{figure}[t!]
\begin{center}
    \includegraphics[width=\linewidth]{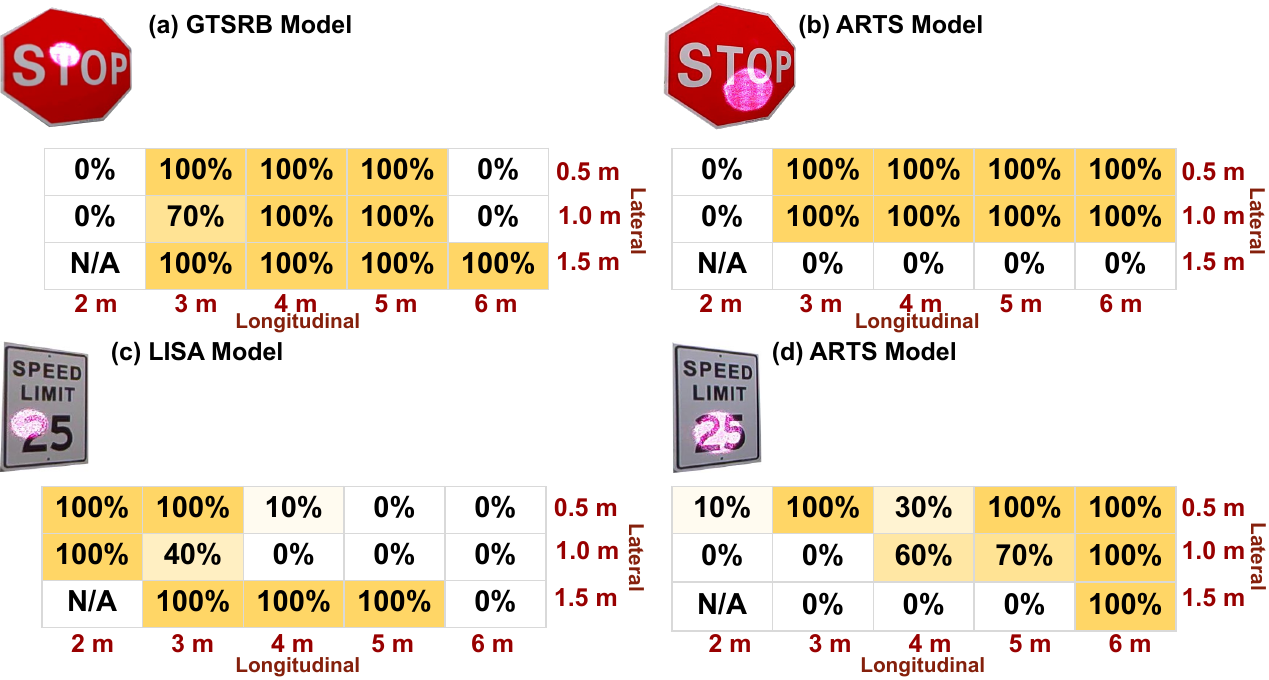}
\end{center}
\caption{SCR for two-stage architecture model with 14 different camera positions. N/A: the traffic sign is out of FOV.
}
\label{fig:robust_camera_scr}
\end{figure}

\noindent\textbf{Robustness to Different Camera Positions.} \label{sec:attack_robust_pos}
Fig.~\ref{fig:robust_camera_asr} and~\ref{fig:robust_camera_scr} show the ASR and SCR of the ILR attack at increasing longitudinal ($d_{vs}$) and lateral ($d_{av}$) distances of the victim camera. The attack is optimized with the traces collected at $d_{vs}$ = 2 m and $d_{av}$ = 1 m and evaluated with real-world experiments at all other victim camera positions. As shown, the lateral direction has a higher impact on attack success than the longitudinal direction. We believe that the ROI cropping and resizing before applying the CNN model inference can cancel the effect of the longitudinal differences, while lateral differences change the viewing angle of the traffic signs,  significantly altering the speckles in the resulting images. As the attack is not optimized for the viewing angle, attack performance is degraded despite applying EoT techniques (See \S\ref{sec:attack_opt}).
Nevertheless, the ASRs remain high, particularly within 1 m lateral translations. Since a road lane is approximately 3.0-3.6 meters wide in real-world scenarios, degradation from different camera viewing angles does not have a major impact on attack performance. For SCR, the stop sign typically has higher values than the speed limit sign, while the speed limit sign has higher ASRs. 

\begin{table}[t!]
\vspace{0.2in}
\centering
\footnotesize
\setlength{\tabcolsep}{5pt}
\setlength{\aboverulesep}{0pt}
\setlength{\belowrulesep}{0pt}
\renewcommand{\arraystretch}{1.2}
\caption{Attack robustness to different angles between the laser emitter and the targeted traffic sign.}
\scalebox{0.75}[1]{
\begin{tabular}{ccccccccccccc}
\toprule
 &  & \multicolumn{2}{c}{Left-20°} &  & \multicolumn{2}{c}{Left-10°} &  & \multicolumn{2}{c}{Right-10°} & & \multicolumn{2}{c}{Right-20°}\\ \cline{3-4} \cline{6-7} \cline{9-10} \cline{12-13}
                             &             & ASR   & SCR   &  & ASR   & SCR   &  & ASR   & SCR  & & ASR & SCR \\ \hline\hline
\multirow{2}{*}{\rotatebox{0}{\begin{tabular}[c]{@{}c@{}}Stop \\ Sign\end{tabular}}}   & GTSRB  & \textbf{100\%} & \textbf{100\%} &  & \textbf{100\%} & \textbf{100\%} &  & \textbf{100\%} & \textbf{100\%} & & \textbf{100\%} & \textbf{100\%} \\ \cline{2-13} 
                             &  ARTS & 50\% & 50\%  &  & \textbf{100\%}  & \textbf{100\%}  &  & \textbf{100\%} & \textbf{100\%} & & 70\% & 70\%\\ \hline\hline
\multirow{2}{*}{\rotatebox{0}{\begin{tabular}[c]{@{}c@{}}Speed\\Limit\end{tabular}}} & LISA  & \textbf{100\%} & \textbf{100\%} &  & \textbf{100\%} & \textbf{100\%}   &  & \textbf{100\%} & \textbf{100\%} & & \textbf{100\%} & 0\%\\ \cline{2-13} 
                             & ARTS  & \textbf{100\%} & \textbf{100\%} &  & \textbf{100\%} & \textbf{100\%} &  & \textbf{100\%} & \textbf{100\%} & & \textbf{100\%} & \textbf{100\%}\\ \bottomrule
\end{tabular}}
\label{tbl:angleRobustness}
\end{table}

\noindent\textbf{Robustness to Different Laser Projection Angles.} \label{sec:attack_robust_las_pos}
Table~\ref{tbl:angleRobustness} lists the ASR and SCR for the ILR attack against the second-stage classifiers at different angles between the laser emitter and the targeted traffic sign. The attack is generated with the laser emitter in front of the target traffic sign and evaluated for four different laser projection angles, spanning a total of 40$^{\circ}$ ($\pm20^{\circ}$ relative to the plane of the traffic sign). As shown, the attack on the stop sign in the GTSRB model has high robustness while for the ARTS model, the ASR drops at 20$^{\circ}$ in both directions. The attack on the speed limit shows a 100\% ASR for all projection angles for both models. The slight performance degradation for the stop sign in ARTS is consistent with the IR speckle pattern variations observed in the camera position and ambient light experiments.

\noindent\textbf{Robustness to Inaccuracy in First-Stage Object Detection.}
We evaluate robustness against first-stage architecture inaccuracies, which can modify the ROI cropping, and consequently alter the input to the second-stage classification model.  We found that the ILR attack is robust against displacement errors of $\leq$8\%. More detailed results are in Appendix~\ref{appendix:robust_eval}.

\nsubsection{Attack Transferability Evaluation}  \label{sec:attack_transfer}

In this section, we evaluate ILR attack generality for different classifiers in the two-stage approach, and for different object detectors in the single- and two-stage approaches. We follow the same experimental setup as in~\S\ref{sec:attack_effectiveness}.
    
\noindent\textbf{Transferability to Different Model Architectures.} \label{sec:transfer_arch}
Table~\ref{tbl:diff_arch} lists the ASR of transferred attacks generated with the CNN models and applied to 3 types of models: DenseNet121~\cite{huang2017densely}, EfficientNet B0~\cite{tan2019efficientnet}, and ResNet50~\cite{he2016deep}. The speed limit sign has significantly higher attack transferability to other models, as the ASR is always 100\%. 
Meanwhile, the stop sign has a low transferability due to the low contrast between the speckle color and the stop sign surface color (red). However, the transferability on EfficientNet has an ASR of 100\%. We hypothesize that the ILR attack causes perturbations in features to which the model has greater sensitivity than it does for adversarial patch attacks.~\cite{eykholt2018physical, chen2018shapeshifter, zhao2018seeing}.

Our results show that the ILR attack can be transferred from one model to another if the two models rely on the same robustness features to determine their predictions, as with the CNN and EfficientNet, but may fail if the features differ, as exemplified by DenseNet121 and ResNet50.

\begin{table}[t!]
\vspace{0.2in}
\centering
\footnotesize
\setlength{\tabcolsep}{3pt}
\setlength{\aboverulesep}{0pt}
\setlength{\belowrulesep}{0pt}
\caption{ASR of the transfer attacks between different model architectures. The attacks are generated by the source model and evaluated in the transferred model.}
\scalebox{0.895}[1]{
\begin{tabular}{ccc||ccc}
\toprule
                       & \multicolumn{2}{r||}{Source Model} & \multicolumn{3}{c}{Transferred Model}    \\
                       &    Dataset                & CNN          & DenseNet121~\cite{huang2017densely} & EfficientNet B0~\cite{tan2019efficientnet} & ResNet50~\cite{he2016deep} \\ \hline\hline
\multirow{2}{*}{\begin{tabular}[c]{@{}c@{}}Stop \\ Sign\end{tabular}}  & ARTS         & \textbf{100\%}        & 0\%         & 0\%             & 0\%      \\ 
                       & GTSRB        & \textbf{100\%}        & 0\%         & \textbf{100\%}           & 0\%      \\ \hline\hline
\multirow{2}{*}{\begin{tabular}[c]{@{}c@{}}Speed \\ Limit\end{tabular}} & ARTS         & \textbf{100\%}        & \textbf{100\%}       & \textbf{100\%}           & \textbf{100\%}    \\ 
                       & LISA         & \textbf{100\%}        & \textbf{100\%}       & \textbf{100\%}           & \textbf{100\%}    \\ \bottomrule
\end{tabular}}
\label{tbl:diff_arch}
\end{table}

\noindent\textbf{Transferability to Different Training Datasets.} \label{sec:transfer_dataset}
We evaluate ILR attack transferability across three datasets (ARTS GTSRB and LISA) using the same CNN model (CNN model details are listed in Appendix~\ref{appendix:model_arch}).
Our evaluation shows high transferability (100\% ASR) for all the datasets. We believe this is due to the large, manipulated area of traffic signs resulting from the speckle. Compared to the results in~\S\ref{sec:transfer_arch}, the model architecture has a more significant impact on attack transferability than does the training data set since it influences the features used for the classification.

\noindent\textbf{Transferability between Different Object Detectors.} \label{sec:transfer_objdet}
Table~\ref{tbl:diff_det} lists the ASRs for  ILR attacks  transferred between different object detectors. As observed in~\S\ref{sec:transfer_arch} and~\S\ref{sec:transfer_dataset}, the ILR attack shows higher transferability even against object detectors. However, the YOLOv3 model trained on the COCO dataset appears more robust with only a 20\% ASR. 
The model trained on the COCO dataset is used for generic object detection rather than specific for traffic sign recognition. Thus, it only has a single class for traffic signs (the stop sign). For this reason, it becomes harder to alter the legitimate prediction of the stop sign with small IR traces compared to a model trained on several different traffic signs.
This result indicates that the first-stage object detector in the two-stage approach can be robust against ILR attacks while the second-stage classifier is still vulnerable. We will discuss it in~\S\ref{sec:discussion}.

\begin{table*}[t!]
\centering
\footnotesize
\setlength{\tabcolsep}{5.5pt}
\setlength{\aboverulesep}{0pt}
\setlength{\belowrulesep}{0pt}
\renewcommand{\arraystretch}{1.1}
\caption{Transfered ILR attack success rates (ASRs) for different object detectors.}
\scalebox{0.895}[1]{
\begin{tabular}{cccccccc}
\toprule
 &               & \multicolumn{6}{c}{Target model}                                                                               \\ \cline{3-8} 
 &               & \multicolumn{2}{c}{Stop}           &  &                         & \multicolumn{2}{c}{Speed}                    \\ \cline{3-4} \cline{7-8} 
 &               &  Faster R-CNN (ARTS) & YOLOv3 (COCO) &  &                         & Faster R-CNN (ARTS) & Faster R-CNN (Mapillary) \\ \cline{1-4} \cline{6-8} 
\multirow{2}{*}{\begin{tabular}[c]{@{}c@{}}Source\\ Model\end{tabular}} & Faster R-CNN (ARTS) & \textbf{100\%} & 20\% & \multirow{2}{*}{} & Faster R-CNN (ARTS) & \textbf{100\%} & \textbf{100\%} \\
 & YOLOv3 (COCO) & \textbf{100\%}              & \textbf{100\%}         &  & Faster R-CNN (Mapillary) & \textbf{100\%}              & \textbf{100\%}                   \\ \bottomrule
\end{tabular}}
\label{tbl:diff_det}
\end{table*}

\nsubsection{Outdoor Evaluation}  \label{sec:eval_outdoor}

\begin{figure}[t!]
\begin{center}
    \includegraphics[width=\linewidth]{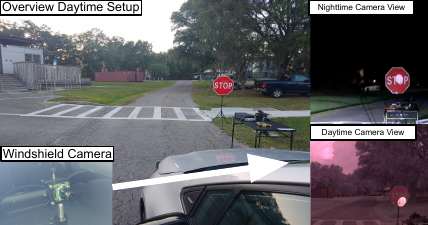}
\end{center}
\caption{Overview of the outdoor experimental scenarios. The setup is used in the daytime (left). The camera view during the attack is at nighttime (Right-top) and daytime (Right-bottom).
}
\label{fig:outdoor_overview}
\end{figure}

To study the effectiveness of the ILR attack in realistic scenarios, we evaluate the attack against the second-stage classification models in a controlled outdoor scenario with the setup described in~\S\ref{sec:attack_effectiveness} under different ambient light conditions (e.g., day and night). Fig.~\ref{fig:outdoor_overview} shows an overview of the evaluation scenario and victim camera view during day and night natural light conditions. We evaluate the two automotive camera sensors: OnSemi and OmniVision. The results of the OmniVision are detailed in Appendix~\S\ref{appendix:dynamic_leopard2}.

\subsubsection{Static Scenarios}  \label{sec:eval_static_outdoor}
We follow the same experimental setup as in ~\S\ref{sec:attack_effectiveness} and perform 10  trials for each experiment.

\noindent\textbf{Nighttime Attack.} We collect attack traces for optimization with the victim camera placed at a 5 m longitudinal distance and a 1 m lateral distance, and then perform the optimized attack in the real world. Similarly to~\S\ref{sec:attack_effectiveness}, we set $d_{as}$ to 3 m. We measure an average ambient light of 120 lux.

For GTSRB stop sign recognition, we achieve 100\% ASR and SCR using 45 mW of power and an average trace diameter of 23 cm, equivalent to 7\% of the entire traffic sign surface. 
For LISA, as used for speed limit signs, we achieve 100\% ASR and 20\% SCR using 46 mW of power, covering 17\% of the traffic sign. 
Finally, we achieve a 100\% ASR and SCR for ARTS on the stop sign. For the speed limit sign, the attack causes a 100\% ASR and a 0\% SCR with a laser power of 115 mW and an average trace diameter of 28 cm, covering 10.6\% of the stop sign. 
We observe that ILR requires higher power compared to the indoor setting because of the different outdoor illuminance compared with artificial light. We believe the degradation of SCR is due to illuminance instability, which we also notice in all our outdoor experiments. 

\noindent\textbf{Daytime Attack.}
During the day, we measured an average ambient light of 982 lux. In this case, we set a shorter distance $d_{as}$ = 1.5 m to reach the required power and pattern size of our optimization methodology for safety constraints. 

For ARTS used for stop sign recognition, we achieve a 100\% ASR and SCR, using a power of 226 mW and 31 cm trace diameter, equivalent to 13.1\% of the stop sign surface. 
For the speed limit sign, we achieve a 100\% ASR for both ARTS and LISA, using a power of 52 mW and an average trace diameter of 17.5 cm, covering an average of 13\% of the speed limit sign. The SCR for ARTS and LISA on the speed limit sign are 50\% and 90\%, respectively.
Finally, for GTSRB,  we achieve an ASR of 100\% and an SCR of 80\% on speed limit with a laser power of 115 mW and an average trace diameter of 31 cm, covering 7.9\% of the traffic sign surface. 

\nsubsubsection{Dynamic Driving Scenarios}\label{sec:outtdoor_dynamic}

We collect the attack traces using the same setup as in the static scenarios (\S\ref{sec:eval_static_outdoor}).
We recorded videos with the victim camera placed in a car moving towards the traffic sign (from 12 meters away from the traffic sign) at three increasingly high speeds: 5, 8, and 13 km/h (approximately 3, 5, and 8 mph)\footnote{We did not evaluate at higher speeds due to the safety and spatial constraints of our testing facility. The demo videos of our experiments are available at \textcolor{blue}{\url{https://sites.google.com/view/cav-sec/ilr-attack}}}.

In this case, the ASR is calculated as the percentage of successful misclassification in terms of the number of successful frames among all the frames collected by the camera. 

\noindent\textbf{Results.}
Table~\ref{tbl:outdoor} shows the ASR in the outdoor driving scenarios for the OnSemi camera. The results are consistent with the indoor robustness experiments in~\S\ref{sec:attack_robust_pos}, i.e., the ILR attack achieves an ASR $>$99\% for all the tested speeds on ARTS and LISA models. For the stop sign, on the other hand, we observe an ASR $>$90\% for ARTS and $>$80.5\% in GTSRB at all speeds. 
The ASR for the ARTS detection model is 100\% in all scenarios. The low SCR for attacks on speed limit sign classification is due to the high sensitivity of the models for the speed limit classification compared to stop sign classification.

Appendix~\S\ref{appendix:dynamic_leopard2} Table~\ref{tbl:outdoor_leopard2} shows results for the OmniVision camera. These results show that our attack achieves high effectiveness in outdoor, moving scenarios, especially in night driving conditions.

\begin{table}[t!]
\vspace{0.2in}
\centering
\footnotesize
\setlength{\tabcolsep}{1.4pt}
\setlength{\aboverulesep}{0pt}
\setlength{\belowrulesep}{0pt}
\renewcommand{\arraystretch}{1.1}
\caption{ASR of the OnSemi camera in the outdoor driving scenarios.}
\scalebox{1.15}[1.1]{
\begin{tabular}{cccccccccccc}
\toprule
                & \multicolumn{4}{c}{Stop Sign}                       &      &  & \multicolumn{4}{c}{Speed Limit}                        &      \\ \cline{2-5} \cline{8-11}
 & \multicolumn{2}{c}{ARTS} & \multicolumn{2}{c}{GTSRB} &  &  & \multicolumn{2}{c}{ARTS} & \multicolumn{2}{c}{LISA} &  \\ \cline{1-6} \cline{8-12} 
           \multicolumn{1}{c|}{Speed}     & ASR & SCR & ASR  & SCR                      & &  & ASR & SCR  & ASR  & SCR  &  \\ \cline{1-12}
           \multicolumn{12}{c}{\textbf{Night Scenario}}\\ \cline{1-12}
\multicolumn{1}{c|}{5 km/h} & \textbf{100\%}  & \textbf{100\%} & 99\% & 90\% &  &  & \textbf{100\%}  & 0\% & 99\%   &  31\% &  \\
\multicolumn{1}{c|}{8 km/h}     & \textbf{100\%}  & \textbf{100\%} & 92\%  & 91\% &  &  & \textbf{100\%}  & 0\%   & \textbf{100\%}   &  0\% &   \\
\multicolumn{1}{c|}{13 km/h}   & \textbf{100\%}   & \textbf{100\%} & 85\%   & 85\%  &  &  & \textbf{100\%}   & 0\%    & 99\%   &  16\% &
 \\ \cline{1-12}
           \multicolumn{12}{c}{\textbf{Day Scenario}}\\ \cline{1-12}
\multicolumn{1}{c|}{5 km/h} & 98\%  & 82\% & 85\% & 57\% &  &  & \textbf{100\%}  & 18\% & \textbf{100\%}   &  98\% &  \\
\multicolumn{1}{c|}{8 km/h}     & \textbf{100\%} & 88\% & 88\%  & 46\% &  &  & \textbf{100\%}  & 50\%   & \textbf{100\%}   &  87\% &   \\
\multicolumn{1}{c|}{13 km/h}   &  91\%  & 75\% & 80\%   & 40\%  &  &  & \textbf{100\%}   & 58\%    & \textbf{100\%}   &  98\% &
 \\ \bottomrule
\end{tabular}}
\label{tbl:outdoor}
\vspace{-0.1in}
\end{table}

\nsection{Defense Evaluation} \label{sec:defense}

In this section, we evaluate existing defenses against patch attacks on ILR attacks and propose a new defense strategy. 

\subsection{Evaluation of generic defenses against patch attacks}  \label{sec:defense_patch}

While invisible to humans, ILR attack traces are visible in camera images.  Thus, existing defense methods against adversarial patch attacks~\cite{eykholt2018physical, zhao2018seeing} are theoretically applicable. So far, two types of defenses against adversarial patch attacks have been proposed: empirical defenses, such as the detection of anomalous patterns in attack patches~\cite{hayes2018visible, gowal2019scalable, yu2021defending, madry2017towards, zhao2018seeing, zhong2022shadows}), and certifiable defenses, which provide theoretical guarantees~\cite{Chiang2020Certified, xiang2021patchguard, levine2020randomized, xiang2022patchcleanser}. As the empirical defenses are known to be generally vulnerable to adaptive attacks~\cite{Chiang2020Certified}, we focus on certifiable defenses.   PatchCleanser~\cite{xiang2022patchcleanser} is the current state-of-the-art for defending classifiers against adversarial patch attacks. 

\vspace{0.05in}
\noindent\textbf{Experimental Setup.} We evaluate the defense capability of PatchCleanser~\cite{xiang2022patchcleanser} on second-stage classifier models in the two-stage architecture.  This architecture scales to a large number of classes and is thus applicable to a more general set of CAV systems. We use the same models as in~\S\ref{sec:attack_effectiveness} -- CNN models trained on the ARTS, GTSRB, and LISA datasets. We generate ILR attacks against 20 scenarios with 2 lateral positions (0.5 m and 1 m) at 2 m from the traffic signs. We limit the diameter of the ILR trace to 12 cm, corresponding to approximately 10 pixels in the image.  In order to focus on defense capabilities, we used our simulated, rather than physical, attack for our evaluation.
PatchCleanser requires the estimated attack patch size as a parameter, thus, we set the patch size to 9 and 12 pixels. The 9-pixel patch is equivalent to the 2\%-pixel patch scenario  in~\cite{xiang2022patchcleanser} (note that this size is smaller than our ILR traces), while the 12-pixel patch (4\%-pixel patch) is designed to cover an ILR trace 10 pixels in diameter.
For the other parameters, we follow the official implementations~\cite{xiang2022patchcleanser}. 
\vspace{0.05in}
\noindent\textbf{Results.} Tables~\ref{tbl:defense_2} and~\ref{tbl:defense_4} (in Appendix~\ref{appendix:patchcleanser_mislabel}), show the defense evaluation of PatchCleanser against the ILR attacks in the 2\%-pixel patch scenario and 4\%-pixel patch scenario, respectively.
The accuracy without any defenses is the accuracy without the PatchCleanser. The clean accuracy is the percentage of correct labels after PatchCleanser is applied.
The certified accuracy is the percentage of correct labels that the PatchCleanser can certify. The mis-certified (false positive) FP is the percentage of \textit{incorrect} labels PatchCleanser certifies.

As shown, our results indicate that PatchCleanser either does not benefit, or decreases, model performance for traffic sign recognition. For example, ILR attacks are not successful (100\% accuracy without PatchCleanser) against the ARTS model on the stop sign because we limit the trace size. 
Nevertheless, PatchCleanser cannot handle them correctly as the clean accuracy is 0\%. As shown in the \textbf{bold} numbers in Table~\ref{tbl:defense_2} and~\ref{tbl:defense_4}, PatchCleanser degrades accuracy by an average of $\geq$12\% for benign cases and 25\% for attack cases for ILR attacks. We believe this is because PatchCleanser's main assumption does not apply to traffic sign recognition. PatchCleanser states that \textit{``model predictions on images without adversarial pixels are generally correct and invariant to the masking operation.''}
This consideration generally holds for image classification tasks whose classes tend to be inferable even if some parts of the image are missing, such as the ImageNet~\cite{deng2009imagenet} and CIFAR-10~\cite{krizhevsky2009learning} datasets. However, traffic sign recognition is an exception to this intuition. For example, a 35 mph sign could be classified as an 85 mph sign if the left side of ``3'' is altered~\cite{tesla_85mph}. Fig.~\ref{fig:defense_example} in Appendix~\ref{appendix:patchcleanser_mislabel} shows examples of PatchCleanser mis-certifying examples of two-rounded, masked images.  A two-round mask hides important text on the traffic sign and causes misclassification in all 36 combinations.  Thus PatchCleanser's agreement-based defense strategy fails in those cases.
Our evaluation shows that while certifiable defenses are typically considered more effective than empirical defenses~\cite{Chiang2020Certified}, this does not mean that they are effectively applicable in every domain. More details are discussed in Appendix~\ref{appendix:patchcleanser_mislabel}.

\cut{
\begin{table}[t!]
\vspace{0.1in}
\centering
\footnotesize
\setlength{\tabcolsep}{2.6pt}
\setlength{\aboverulesep}{0pt}
\setlength{\belowrulesep}{0pt}
\renewcommand{\arraystretch}{1.1}
\caption{Defense evaluation of PatchCleanser against the ILR attacks. The certified TP is the rate of correct labels that PatchCleanser can certify. The miscertified FP is the rate of \textit{incorrect} labels but PatchCleanser certifies.}
\scalebox{0.89}[1]{
\begin{tabular}{cccccccccc}
\toprule
             & \multicolumn{4}{c}{Benign}  &  & \multicolumn{4}{c}{Attack} \\ \cline{2-5} \cline{7-10} 
 & \multicolumn{2}{c}{Stop Sign} & \multicolumn{2}{c}{Speed Limit} &  & \multicolumn{2}{c}{Stop Sign} & \multicolumn{2}{c}{Speed Limit} \\ \cline{2-5} \cline{7-10} 
             & GTSRB & ARTS & LISA  & ARTS &  & GTSRB & ARTS & LISA & ARTS \\ \midrule
No Defense Acc.$\uparrow$   & 93\%  & 93\% & 100\% & 93\% &  & 7\%   & 93\% & 0\%  & 0\%  \\
Clean Acc.$\uparrow$          & \textbf{86\%}  & \textbf{43\%} & \textbf{64\%}  & \textbf{71\%} &  & 7\%   & \textbf{36\%} & 0\%  & 0\%  \\
Certified TP$\uparrow$ & 0\%   & 0\%  & 0\%   & 0\%  &  & 0\%   & 0\%  & 0\%  & 0\%  \\
Miscertified FP$\downarrow$ & \underline{7\%}  & \underline{50\%} & \underline{36\%} & \underline{21\%}  &  & \underline{36\%}  & \underline{57\%} & \underline{64\%} & \underline{14\%} \\ \bottomrule
\end{tabular}}
\label{tbl:defense}
\end{table}
}

\begin{table}[t!]
\vspace{0.2in}
\centering
\footnotesize
\setlength{\tabcolsep}{0.8pt}
\setlength{\aboverulesep}{0pt}
\setlength{\belowrulesep}{0pt}
\renewcommand{\arraystretch}{1.1}
\caption{Defense evaluation of PatchCleanser against the ILR attacks with the \textit{2\%-pixel patch}, as used in the original PatchCleanser paper~\cite{xiang2022patchcleanser}. The certified TP is the rate of correct labels that PatchCleanser can certify. The mis-certified FP is the rate of \textit{incorrect} labels  PatchCleanser certifies.}
\scalebox{0.895}[1]{
\begin{tabular}{cccccccccccc}
\toprule
                & \multicolumn{4}{c}{Benign}                       &      &  & \multicolumn{4}{c}{Attack}                        &      \\ \cline{2-5} \cline{8-11}
 & \multicolumn{2}{c}{Stop Sign} & \multicolumn{2}{c}{Speed Limit} &  &  & \multicolumn{2}{c}{Stop Sign} & \multicolumn{2}{c}{Speed Limit} &  \\ \cline{2-6} \cline{8-12} 
                & GTSRB & ARTS & LISA  & ARTS                      & Avg. &  & GTSRB & ARTS  & LISA  & ARTS                      & Avg. \\ \cline{1-6} \cline{8-12} 
No Defense Acc.$\uparrow$ & 93\%  & 93\% & 100\% & \multicolumn{1}{c|}{93\%} & 95\% &  & 15\%  & 100\% & 0\%   & \multicolumn{1}{c|}{0\%}  & 29\% \\
Clean Acc.$\uparrow$      & 93\%  & 93\% & \textbf{71\%}  & \multicolumn{1}{c|}{\textbf{71\%}} & \textbf{82\%} &  & 15\%  & \textbf{0\%}     & 0\%   & \multicolumn{1}{c|}{0\%}  & \textbf{4\%}  \\
Certified Acc.$\uparrow$    & 0\%   & 64\% & 0\%   & \multicolumn{1}{c|}{0\%}  & 16\% &  & 0\%   & 0\%     & 0\%   & \multicolumn{1}{c|}{0\%}  & 0\%  \\
Miscertified FP$\downarrow$ & \underline{0\%}   & \underline{0\%}  & \underline{29\%}  & \multicolumn{1}{c|}{\underline{21\%}} & \underline{13\%} &  & \underline{5\%}   & \underline{90\%}  & \underline{100\%} & \multicolumn{1}{c|}{\underline{20\%}} & \underline{54\%} \\ \bottomrule
\end{tabular}}
\label{tbl:defense_2}
\vspace{-0.1in}
\end{table}

\subsection{Proposed Alternative Defense Strategies} 
\label{sec:defense_detection}
While certifiable defenses are not sufficient to eliminate ILR, and applying optical IR filters defeats the advantage of using the IR light components to detect obstacles, alternative strategies can be adopted to evaluate the trustworthiness of traffic sign recognition. We propose a detection strategy based on the physics-based characteristics of laser light reflections. 

\begin{figure}[t!]
\begin{center}
    \includegraphics[width=\linewidth]{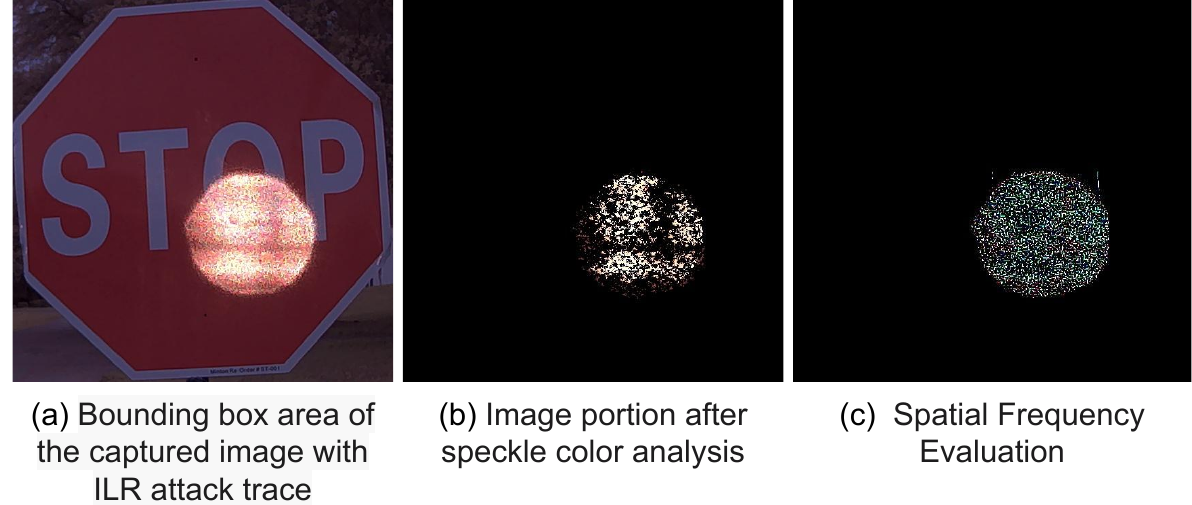}
\end{center}
\caption{Speckle Detection on a stop sign. (a) Real-world attack trace during daytime. (b) Color masking result. (c) Spatial Frequency analysis of the selected portion.}
\label{fig:proposed_defense}
\end{figure}

\noindent\textbf{Speckle Features.} As described in~\S\ref{sec:ir_feasibility}, laser beam light generates speckle patterns when coherent light diffuses off of a rough surface (such as that of a traffic sign), causing interference. The resulting reflected speckle pattern appears as a random distribution of bright and dark spots in images.  The pattern varies based on speckle location, surface roughness, camera settings, captured images pixel resolution, and the optical power of the reflection, as shown in~\ref{fig:ilr_attack_overview}.

The spatial pixel frequency of an image refers to the rate of change in intensity values from one pixel to its neighboring pixels. Laser speckles exhibit high spatial frequency, indicating that adjacent pixels have significant variations in intensity in smaller spaces~\cite{goodman1976some}. Additionally, the speckle can only manifest as a monotonous false color, such as magenta, purple, or orange, depending on image sensor type, the IR light, and the ambient light condition described in~\S\ref{ir_color}. Leveraging these unique features~\cite {goodman1975statistical, yeh2012robust}, various defense strategies can be adopted to identify and locate ILR attack traces within images, independently of speckle size and shape. 

\noindent\textbf{Color-Frequency Detection.}
Taking inspiration from image processing techniques~\cite{shapley1985spatial, srinivasan2008statistical, gonzales1987digital}, a detection methodology can begin by extracting salient regions from captured traffic sign images that may contain false colors as a result of an ILR attack, A spatial frequency analysis is then performed to determine whether selected regions also manifest higher spatial frequencies, indicating a potential attack.
For instance, for the OnSemi camera, our outdoor experiments show that at high ambient illuminance, the speckle color falls within the range of \texttt{\#FFB266} to \texttt{\#CC6600}, while during low ambient light, it ranges between \texttt{\#CF9FFF} and \texttt{\#DA70D6}. These color ranges can be extracted from the camera output and used as references for the subsequent frequency analysis. 

\noindent\textbf{Proof-of-Concept Analysis.} 
To test the feasibility of the approach, as a proof of concept, we implement the strategy on a random selection of real-world images collected with the OnSemi camera from all of the outdoor attack scenarios tested in~\S\ref{sec:eval_outdoor}.
We build two sample datasets containing both speed limit and stop signs, collected during daytime and nighttime respectively. Each dataset consists of 75 benign samples and 75 attack samples. 
We then extract from the ARTS model output the traffic sign bounding box areas to perform the analysis.
To apply the methodology, we first use a color masking technique, with potential IR light color ranges measured empirically. Next, to differentiate IR speckles from naturally occurring image components, we apply a Gaussian smoothing low-pass filter~\cite{gonzales1987digital} to separate the low-frequency components of the image, and to retain regions with high-frequency components, as shown in Figure~\ref{fig:proposed_defense}. We then identify a potential ILR attack if more than 1\% of an extracted region exhibits high spatial frequencies. This threshold is chosen because benign traffic sign images typically contain few high spatial frequency components. This preliminary test achieves a 96\% TPR and a 2.7\% FPR for the daytime data. For the nighttime data instead, the methodology shows 92\% TPR and 6.7\% FPR.
More sophisticated techniques such as color segmentation~\cite{cheng2001color} and cross-correlation~\cite{wang2013fast, yang2010fast} can also be used for detecting known patterns. 
\section{Discussion and Limitations} \label{sec:discussion}

\noindent\textbf{CAV Safety Implications.} As demonstrated in~\S\ref{sec:outtdoor_dynamic}, the ILR attack can achieve a nearly 100\% ASR in outdoor driving scenarios, particularly at night.  This can severely undermine CAV safety. Furthermore, we note that the ILR attack can be easily enhanced if the attacker uses multiple laser emitters to affect wider areas or to generate saturated IR speckles. 
Using our methodology, the attacker can alter traffic signs of any size by scaling the power and size of the attack trace proportionally, and of any color by adjusting the traces according to the traffic sign surface color, as shown in~\S\ref{sec:diff_modeling}. Thus, we recommend that CAV companies either use IR filters, or deploy adequate defenses such as the one proposed in~\S\ref{sec:defense_detection}.

\noindent\textbf{Single-Stage v.s. Two-Stage Architectures.}
We observe that both single-stage and two-stage architectures are vulnerable to the ILR attack in~\S\ref{sec:attack_effectiveness} and~\S\ref{sec:attack_robust}. The generic object detector trained with the COCO dataset shows higher robustness to ILR, thus it might appear suitable to use for a single-stage architecture, or the first stage of a two-stage architecture.
However, the single-stage architecture is not applicable to higher SAE autonomy levels ~\cite{SAEAutonomyLevels}, as it does not scale to a large number of traffic sign classes~\cite{ertler2020mapillary}.

\noindent\textbf{Daytime Attack under Strong Sunshine.}
While we confirmed that the ILR attack is robust to some lighting conditions in~\S\ref{sec:attack_robust_light}, the ILR attack is not likely to work in strong sunshine without raising the laser power to Class 4 (above 500mW)~\cite{lasersafety}. We could not evaluate this condition because of safety hazards and the uncontrollability of the sunshine level.
However, our ILR attack should achieve at least equal or higher robustness than the attacks that use incoherent light, such as projection of an image of a person on the ground~\cite{nassi2020phantom} or projection of an adversarial pattern onto a traffic sign~\cite{lovisotto2021slap} as they use incoherent lights not robust to reflection.  

\noindent\textbf{Laser Safety.}
All of our real-world experiments were conducted in closed controlled environments by trained personnel. 
For outdoor experiments, the power of a class 3-B laser was used (see Appendix~\ref{appendix:safety} for the details).
\nsection{Conclusion} \label{sec:conclusion}

We propose ILR, a novel, invisible attack vector that can cause CAV traffic sign recognition systems to misclassify traffic signs.  Our attack uses an IR laser to reflect a pattern onto a target sign that is invisible to humans, but visible to CAV cameras that lack IR filters.  Unlike previous attacks, our attack does not require continuous aiming of a moving CAV.

To maximize attack efficacy, we designed a novel methodology to optimize the attack using image difference-based IR trace modeling and interpolation. We evaluated the effectiveness of our attack against two major traffic sign architectures achieving a 100\% success rate for indoor experiments and $\geq$80.5\%  in outdoor driving scenarios. Finally, we determined that certifiable defenses have limited applicability to the traffic sign recognition domain. Thus we propose an alternative defense technique based on speckle detection.

\section*{Acknowledgements}
 
We thank the anonymous shepherd and reviewers for their
valuable comments. This research was supported in part by the NSF CNS-1932464, CNS-1929771, CNS-2145493, USDOT UTC Grant 69A3552047138, 
JST CREST JPMJCR23M4,
and unrestricted research funds from Toyota InfoTech Labs. We want to thank Himanandhan Reddy Kottur for his help with the outdoor experiments.

{\small
\bibliographystyle{IEEEtran}
\bibliography{main.bib}
}

\appendix
\subsection{IR Laser Power Attenuation}
\label{appendix:laser_power_attenuation}
The laser optical power emitted by the attacker laser module can be given by $P = I \cdot A$, where $I$ is the beam's irradiance (power per unit area) and $A$ is the cross-section area of the laser beam. For an attacker distance $d_{as}$ from the target traffic sign, the irradiance of the beam decreases according to inverse square law as $I = P \big/ (4 \pi \cdot d^2_{as})$. The cross-section area of the beam $A$ at $d_{as}$ can be measured as $A = (\pi \cdot d_{as}^2 \cdot \tan^2 \theta) \big/ 4$, where $\theta$ is the divergence angle, controlled by the attacker using the two lens setup. Thus the resulting laser optical power at the traffic sign surface $P_f$ at $d_{as}$ with divergence angle $\theta$ can be given by $P_f = (P_a \cdot \tan^2 \theta) \big/ (4 \pi)$.

\subsection{Laser Speckle Modeling Details} \label{appendix:empirical}
 To accurately synthesize the pixel distribution of the attack traces, we use image differencing to extract the RGB pixel intensity variation in our controlled closed indoor scenario (we follow a similar procedure for outdoor scenarios except for the illuminance setting). 

For the indoor setting, we use the same attack setup described in \S\ref{sec:attack_setup} and we locate the victim camera at a 0.5 m $d_{av}$ from the IR emitter. We set $d_{as}$ to 3 m and the room ambient light to 100 Lux. We then capture the images of a stop sign without and during the attack and extract the pixel intensity variation caused by the IR patterns in the form of RGB pixel difference between both images. 
These intensity differences are then applied on to the benign traffic sign image to simulate IR beams targeted at different sign locations for optimization as shown in Fig.~\ref{fig:simulation_demo}. 

We account for the temporal image noise in the target camera by averaging ten consecutive frames for each benign and attack case. Further, we observe that the RGB pixel offset values depend on the camera’s perceived surface color for the selected traffic sign. To model this, we first collect the attack traces with a baseline traffic sign surface color (e.g., the RGB pixel values that correspond to the red color of the stop sign). We then synthesize the IR pattern on the target traffic sign surface colors (different from the baseline), by measuring the average offset in the RGB values between the baseline and the target traffic sign color.

\subsection{Pixel-wise Spline-based Interpolation} \label{appendix:spline}

As a baseline method for the trace image interpolation, we design the pixel-wise spline-based interpolation in which we simply apply the cubic spline interpolation for each pixel.
This method consists of three steps, (1) Spline fitting: The real-world traces are synthesized by a pixel-wise cubic interpolation function to model the RGB pixel distribution changes for each trace individually; (2) Spline interpolation: For a desired emitted laser power, two adjacent real-world traces are used to calculate the RGB pixel values of the trace; and (3) using weighted averages between the real-world traces and the traces generated in step (2), the RGB pixel values are derived at the desired diameter.

\subsection{CNN Model Architecture} \label{appendix:model_arch}

Table~\ref{tbl:model_arch} lists the architecture of the CNN model we use. The input image size is 60$\times$60 pixels. This model achieves one of the highest performances on the GTSRB dataset~\cite{cnn_gtsrb}. 

\begin{table}[htb]
\vspace{0.1in}
\small
\centering
\caption{CNN model architecture.}
\label{tbl:model_arch}
\begin{tabular}{ll}
\toprule
Layer (type)                        & Output Shape                                         \\ \midrule
0. Input                               &  (batch\_size, 60, 60, 3) \\
1. Conv2D                     & (batch\_size, 28, 28, 16)                            \\
2. Conv2D                  & (batch\_size, 26, 26, 32)                            \\
3. MaxPooling2D       & (batch\_size, 13, 13, 32)                            \\
4. BatchNorm    & (batch\_size, 13, 13, 32)                            \\
5. Conv2D                  & (batch\_size, 11, 11, 64)                            \\
6. Conv2D                  & (batch\_size, 9, 9, 128)                             \\
7. MaxPooling2     & (batch\_size, 4, 4, 128)                             \\
8. BatchNorm & (batch\_size, 4, 4, 128)                             \\
9. Flatten                   & (batch\_size, 18432)          \\
10. Dense                       & (batch\_size, 512)                                   \\
11. BatchNorm & (batch\_size, 512)                                   \\
12. Dropout                   & (batch\_size, 512)                                   \\
 13. Dense                    & (batch\_size, 43)                                    \\ \bottomrule
\end{tabular}
\end{table}

\subsection{Detailed Setup of Random Attacks} 
\label{appendix:random_attack}

A naive attacker might decide to project the IR pattern onto random locations on a sign and observe the behavior of the victim CAV.
To show the consequences of a random attack on both single-stage and two-stage architectures,
we use the setup discussed in~\S\ref{sec:attack_setup} to collect the IR traces from the OnSemi camera~\cite{leopard_camera} for the stop and speed limit signs.  
For this analysis, we set the pattern diameter to $ D = 15$ cm as in our indoor evaluation. To avoid sensor saturation, we used a 51 mW laser power. We then randomly pick a location to place the IR pattern within the traffic signs.
As shown in Table~\ref{tbl:attack_effectivness}, the random attack has an ASR of $\leq$20\% against the stop sign, and $\leq$20\% against the speed limit sign, for the two architectures. We use the random attack as a comparison baseline to demonstrate our optimized ILR attack methodology's effectiveness in~\S\ref{sec:methodology}.

\subsection{Robustness to Inaccuracy in First-Stage Object Detection} 
\label{appendix:robust_eval}

\begin{figure}[t!]
\begin{center}
    \includegraphics[width=\linewidth]{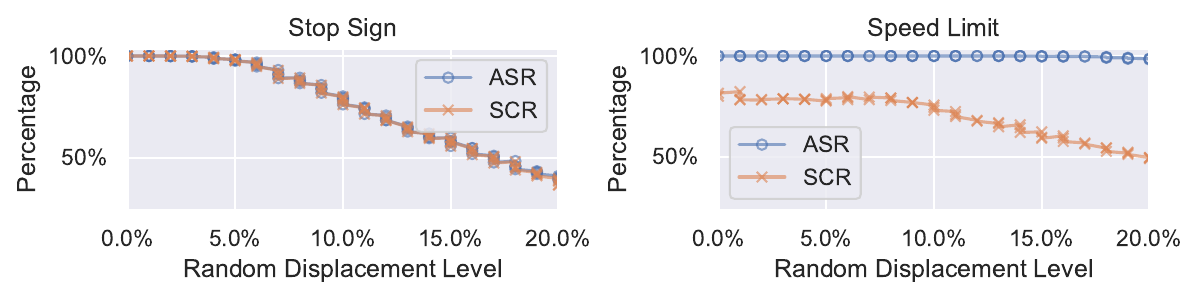}
\end{center}
\vspace{-0.15in}
\caption{Evaluation results of bounding box position noise detected in the first stage.
Given noise level $\delta$, the resulting perturbation, $\mathcal{U}$, obeys: $\mathcal{U}(- \delta, \delta)$ = percentage of  bounding box width or height.  }
\label{fig:robustness_stage1}
\end{figure}

While we focused more on attacking the second-stage classification model for the two-stage architecture, we also evaluated how inaccuracies in the first stage can change the automatic bounding cropping and consequently alter the input of the second-stage classification model. To evaluate the impact of the inaccuracy on the classification results, we apply vertical and horizontal translation noise to our manually annotated bounding boxes. For the stop sign, we use the CNN model trained on the GTSRB dataset. For the speed limit sign, we use the CNN model trained on the LISA dataset.
Fig.~\ref{fig:robustness_stage1} shows the ASR for random vertical and horizontal displacement. Since the bounding box sizes are different for each image, we use the percentage over the width and height of the bounding box as a displacement level, $\delta$, instead of the corresponding sizes in pixels. Given $\delta$, we generate a random number under the uniform distribution $\mathcal{U}(- \delta, \delta)$ and displace the bounding box based on the result. 
For example, a 10-pixel displacement will be applied on a bounding box with 100-pixel height and width if the random number is 10\%.
As shown, bounding box inaccuracy has a greater impact on the stop sign than the speed limit sign. The ASR and SCR for the stop sign decrease with increasing noise levels. In contrast, the ASR for the speed limit sign is always 100\%, while the SCR eventually starts to decrease around a noise level of 8\%. 

We hypothesize that these results are due to the shape, and the resulting pixel RGB values, of the attack beam traces. For example, the majority of attacks against the speed limit signs are classified as stop signs, as shown in Fig.~\ref{fig:simulation_demo}.  This suggests that the beam and stop sign may have similar features, which result in similar classifications.  Thus, small translations of the IR spot can negate attacks, resulting in the correct stop sign classification. 

\subsection{Considerations over PatchCleanser}
\label{appendix:patchcleanser_mislabel}

\begin{table}[h!]
\vspace{0.2in}
\centering
\footnotesize
\setlength{\tabcolsep}{0.8pt}
\setlength{\aboverulesep}{0pt}
\setlength{\belowrulesep}{0pt}
\renewcommand{\arraystretch}{1.1}
\caption{Defense evaluation of PatchCleanser against the ILR attacks with the \textit{4\%-pixel patch}, which can cover the ILR trace. The certified TP is the rate of correct labels that PatchCleanser can certify. The miscertified FP is the rate of \textit{incorrect} labels but PatchCleanser certifies.}
\scalebox{0.895}[1]{
\begin{tabular}{cccccccccccc}
\toprule
                & \multicolumn{4}{c}{Benign}                       &      &  & \multicolumn{4}{c}{Attack}                        &      \\ \cline{2-6} \cline{8-12} 
 & \multicolumn{2}{c}{Stop Sign} & \multicolumn{2}{c}{Speed Limit} &  &  & \multicolumn{2}{c}{Stop Sign} & \multicolumn{2}{c}{Speed Limit} &  \\ \cline{2-5} \cline{8-11}
                & GTSRB & ARTS & LISA  & ARTS                      & Avg. &  & GTSRB & ARTS  & LISA  & ARTS                      & Avg. \\ \cline{1-6} \cline{8-12} 
No Defense Acc.$\uparrow$ & 93\%  & 93\% & 100\% & \multicolumn{1}{c|}{93\%} & 95\% &  & 15\%  & 100\% & 0\%   & \multicolumn{1}{c|}{0\%}  & 29\% \\
Clean Acc.$\uparrow$      & \textbf{86\%}  & \textbf{43\%} & \textbf{64\%}  & \multicolumn{1}{c|}{\textbf{71\%}} & \textbf{66\%} &  & 15\%  & \textbf{0\%}     & 0\%   & \multicolumn{1}{c|}{0\%}  & \textbf{4\%}  \\
Certified Acc.$\uparrow$    & 0\%   & 0\%  & 0\%   & \multicolumn{1}{c|}{0\%}  & 0\%  &  & 0\%   & 0\%     & 0\%   & \multicolumn{1}{c|}{0\%}  & 0\%  \\
Miscertified FP$\downarrow$ & \underline{7\%}   & \underline{50\%} & \underline{36\%}  & \multicolumn{1}{c|}{\underline{21\%}} & \underline{29\%} &  & \underline{0\%}   & \underline{100\%} & \underline{100\%} & \multicolumn{1}{c|}{\underline{85\%}} & \underline{71\%} \\ \bottomrule
\end{tabular}}
\label{tbl:defense_4}
\end{table}

\begin{figure}[h!]
\begin{center}
    \includegraphics[width=\linewidth]{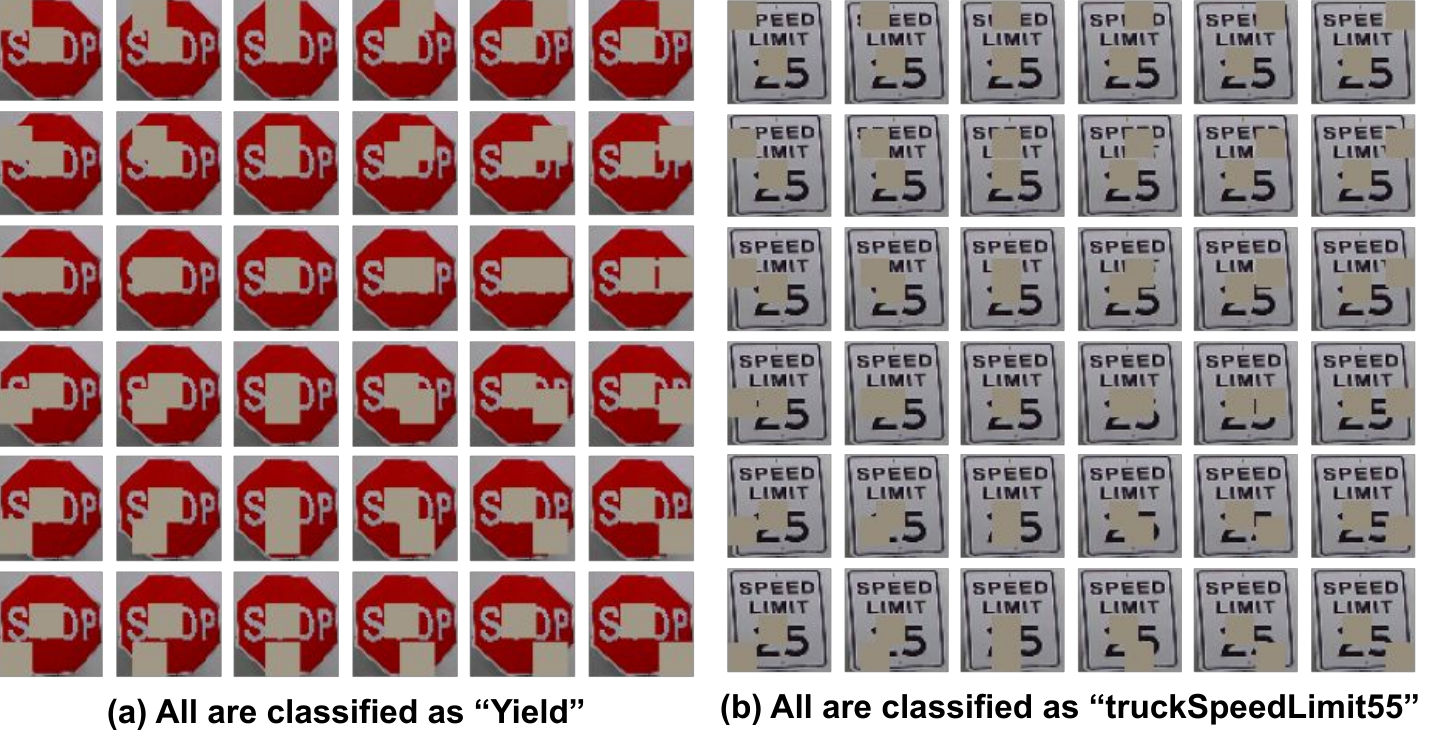}
\end{center}
\caption{Mis-certified examples of two-round masked images. (a) For the stop sign, all images are classified as a ``Yield'' sign in the 4\%-pixel patch scenario. (b) For the speed limit sign, all images are classified as a ``truckSpeedLimit55'' sign in the 2\%-pixel patch scenario.}
\label{fig:defense_example}
\end{figure}

PatchCleanser and PatchGuard  \cite{xiang2021patchguard}, and current state-of-the-art defenses, assume that classification models can return a correct prediction even if a small portion of the image is masked.  However, this assumption does not always hold for traffic sign recognition, since any portion of the sign has the potential to be important for correct classifications.  
As listed in Tables~\ref{tbl:defense_2} and~\ref{tbl:defense_4}, the certified accuracy of PatchCleanser is significantly lower than the reported ($>$60\% certified accuracy) on ImageNet~\cite{deng2009imagenet}. Even for benign cases, the certified accuracies are 16\% for the 2\%-pixel patch scenario and 0\% for the 4\%-pixel patch scenario. In the 2\%-pixel patch scenario, the clean accuracy in the benign case (82\% on average) is close to the reported clean accuracy ($>$80\%) on  ImageNet, but it drops to a 66\%  average in the 4\%-pixel patch scenario. 
For the attack cases, the clean accuracies are even worse (4\%) than the accuracy without PatchCleanser (29\%). Furthermore, PatchCleanser mis-certifies 33.5\% of cases for the 2-pixel patch scenario and 50\% of cases for the 4-pixel patch scenario (averages of the \underline{underlined} numbers in Tables~\ref{tbl:defense_2} and~\ref{tbl:defense_4}) and does not have any correctly certified cases for the 4-pixel patch scenario. This means that the two-round masking of PatchCleanser itself works as an attack, with prediction agreement occurring for a wrong label. Fig.~\ref{fig:defense_example} shows mis-certified examples of the two-round masked images. The two-round mask hides important text on the traffic sign and causes misclassification in all 36 combinations. These images might also be challenging for humans to classify correctly.

Our ILR attack can break another prerequisite for PatchCleanser -- that the attack trace size is known in advance. The size of an ILR attack trace is relative to the target sign size and it can be increased without reducing attack stealthiness  using our methodology described in~\ref{sec:methodology} (note that to maintain a constant trace intensity, the attacker would need to change the laser power based on the distance). 
Additionally, the circular shape of the ILR attack trace cannot be used in PatchCleanser as it is since a significant number of patches are required for circular masks to meet the $\mathcal{R}$-covering conditions.

\subsection{Outdoor Evaluation of the OmniVision Camera}
\label{appendix:dynamic_leopard2}

\begin{table}[h!]
\centering
\footnotesize
\setlength{\tabcolsep}{3.5pt}
\setlength{\aboverulesep}{0pt}
\setlength{\belowrulesep}{0pt}
\renewcommand{\arraystretch}{1.2}
\caption{ASR of ILR attacks on Omnivision in the outdoor static scenarios.}
\scalebox{1}[1]{
\begin{tabular}{cccclcclcc}
\toprule
 &  & \multicolumn{2}{c}{Night.} &  & \multicolumn{2}{c}{Day} \\ \cline{3-4} \cline{6-7} \cline{9-10} 
                             &             & ASR   & SCR   &   & ASR   & SCR   \\ \hline\hline
\multirow{2}{*}{\rotatebox{0}{\begin{tabular}[c]{@{}c@{}}Stop \\ Sign\end{tabular}}}   & ARTS  & 100\% & 100\% &  & 100\% & 20\%  \\ \cline{2-10} 
                             & GTSRB & 100\% & 100\%  &  & 100\% & 90\% &  \\ \hline\hline
\multirow{2}{*}{\rotatebox{0}{\begin{tabular}[c]{@{}c@{}}Speed\\Limit\end{tabular}}} & ARTS  & 100\% & 100\% &   & 100\% & 50\% &  \\ \cline{2-10} 
                             & LISA  & 100\% & 0\% &  & 100\% & 100\% \\ \bottomrule
\end{tabular}}
\label{tbl:outdoor_leopard2_static}
\end{table}

\begin{table}[h!]
\vspace{0.2in}
\centering
\footnotesize
\setlength{\tabcolsep}{1.5pt}
\setlength{\aboverulesep}{0pt}
\setlength{\belowrulesep}{0pt}
\renewcommand{\arraystretch}{1.1}
\caption{ASR of ILR attacks on Omnivision in the outdoor dynamic scenarios.}
\scalebox{1.15}[1.1]{
\begin{tabular}{cccccccccccc}
\toprule
                & \multicolumn{4}{c}{Stop Sign}                       &      &  & \multicolumn{4}{c}{Speed Limit}                        &      \\ \cline{2-5} \cline{8-11}
 & \multicolumn{2}{c}{ARTS} & \multicolumn{2}{c}{GTSRB} &  &  & \multicolumn{2}{c}{ARTS} & \multicolumn{2}{c}{LISA} &  \\ \cline{1-6} \cline{8-12} 
           \multicolumn{1}{c|}{Speed}     & ASR & SCR & ASR  & SCR                      & &  & ASR & SCR  & ASR  & SCR  &  \\ \cline{1-12}
           \multicolumn{12}{c}{\textbf{Night Scenario}}\\ \cline{1-12}
\multicolumn{1}{c|}{5 km/h} & 100\%  & 95\% & 100\% & 92\% &  &  & 95\%  & 0\% & 100\%   &  0\% &  \\
\multicolumn{1}{c|}{8 km/h}     & 100\%  & 78\% & 100\%  & 85\% &  &  & 89\%  & 0\%   & 100\%   &  6\% &   \\
\multicolumn{1}{c|}{13 km/h}   & 100\%   & 85\% & 100\%   & 90\%  &  &  & 96\%   & 0\%    & 100\%   &  1\% &
 \\ \cline{1-12}
           \multicolumn{12}{c}{\textbf{Day Scenario}}\\ \cline{1-12}
\multicolumn{1}{c|}{5 km/h} & 100\%  & 54\% & 99\% & 39\% &  &  & 100\%  & 18\% & 100\%   &  96\% &  \\
\multicolumn{1}{c|}{8 km/h}     & 100\% & 10\% & 99\%  & 94\% &  &  & 100\%  & 50\%   & 100\%   &  100\% &   \\
\multicolumn{1}{c|}{13 km/h}   &  100\%  & 11\% & 100\%   & 80\%  &  &  & 100\%   & 58\%    & 100\%   &  100\% &
 \\ \bottomrule
\end{tabular}}
\label{tbl:outdoor_leopard2}
\end{table}

\subsection{Considerations on Laser Safety}
\label{appendix:safety}
In our experiments, we use the maximum emission power of 80mW in controlled indoor scenarios and 115mW in outdoor controlled scenario (daytime), below the  3-B class laser limit (= 500mW)~\cite{lasersafety}.
The maximum permissible exposure (MPE)~\cite{mpe_formula} of a 780 nm continuous class 3-B laser with an exposure time $t > 10$ seconds is given by
$MPE = 10^{2 \cdot (w - 0.7)} \cdot 10^{-3}$,
where $w$ is the wavelength of the IR laser. Using this equation, at $w = 780$ nm, $MPE = 0.33$~mW/cm$^2$. As an example, for a given 45 mW optical power, the emitter energy is equivalent to 57.7 mW/cm$^2$, nearly 175 times more than the MPE. However, the IR beam's energy can be reduced to below the MPE values by increasing the beam diameter to 3.6 times the original size (1.3 cm for our setup). From this analysis, an IR pattern diameter of nearly 5 cm is required in order to follow MPE guidelines.  Using our ILR attack configuration, which considers a diverging beam, the resulting IR pattern diameter at 45 mW is 17 cm. 

\end{document}